\documentclass[
twocolumn,
superscriptaddress,
showpacs,
nofootinbib,
nobibnotes,
amsmath,amssymb,
aps,
pra,
floatfix,
]{revtex4}

\usepackage{graphicx}
\usepackage{epstopdf}
\usepackage{dcolumn}
\usepackage{bm}
\usepackage{natbib}
\usepackage[caption=false]{subfig}

\def \be {\begin{equation}}
\def \ee {\end{equation}}

\begin{document}

\preprint{APS/123-QED}

%
\title{Optical bistability in one dimensional doped photonic crystals \\ with spontaneously generated coherence}
%
\author{Shahnaz Aas}
\affiliation{Department of Physics, Bilkent University, 06800, Bilkent, Ankara, Turkey}
\author{\"{O}zg\"{u}r E. M\"{u}stecapl{\i}o\u{g}lu}
\email{omustecap@ku.edu.tr}
\affiliation{Department of Physics, Ko\c{c} University, \.Istanbul, 34450, Turkey}

\date{\today}
\begin{abstract}
We investigate optical bistability in a multilayer one-dimensional photonic crystal where the central layer is doped with $\Lambda$-type 
three level atoms. We take into account the influence of spontaneously generated coherence when the lower atomic levels are sufficiently close to each other, in which case Kerr-type nonlinear response of the atoms is enhanced. We calculate the propagation of a probe beam 
in the defect mode window using numerical nonlinear transfer matrix method. We find that Rabi frequency of a control field acting on the defect layer and the detuning
of the probe field from the atomic resonance can be used to control the size and contrast of the hysteresis loop and the threshold of the optical bistability. In particular we find that, at the optimal spontaneously generated coherence, three orders of magnitude lower threshold can be achieved relative to the case without the coherence.
\end{abstract}




\maketitle
\section{\label{sec:intro}Introduction}
Optical bistability (OB) is a striking manifestation of nonlinear behavior in an optical system where
two separate stationary output states are possible for a given input~\cite{gibbs_optical_1985,abraham_optical_1982}.
Technologically, this fundamental nonlinear phenomenon is used for all-optical logic and memory
operations~\cite{soljacic_optimal_2002,barclay_nonlinear_2005}. 
Modern applications demand more compact, more noise tolerant, and faster OB based
devices that can operate under lower power thresholds over a wide range of multistability.

Nonlinear photonic crystal (PC) systems~\cite{Scalora_optical_1994,Danckaert_dispersive_1991,Agranovich_optical_1991}
are microphotonic devices, such as photonic switches~\cite{mingaleev_nonlinear_2002,yanik_high-contrast_2003,mingaleev_all-optical_2006}, diodes~\cite{tocci_thinfilm_1995,zhao_design_2006,xue_highly_2010}, or 
transistors~\cite{soljacic_nonlinear_2003,yanik_all-optical_2003}, which offer sub-picosecond
operation times at miliwatt power levels and suitability for large scale
optical integration~\cite{bravo-abad_enhanced_2007}. 
They can be tailored for efficient optical switching by using embedded atoms in PCs~\cite{guo_controllable_2009,john_optical_1996,john_collective_1997,ma_optical_2011,takeda_self-consistent_2011, vujic_coherent_2007}. Conventional OB in PCs utilizes dynamic shifting of the band edge while the doped PCs allow 
for dispersive OB via dynamic shifting of the defect mode~\cite{wang_dispersive_1997}. One-dimensional multilayer PC (1DPC) systems~\cite{Lidorikis_optical_1997,wang_dispersive_1997,novitsky_bistable_2008} 
are considered for controlling OB. Adding extra coating layers~\cite{gupta_dispersive_1987}, phase matching layer~\cite{he_combined_1992}, negative index layer~\cite{jose_controlling_2009}, or subwavelenght layers~\cite{hou_transmission_2008} next to the nonlinear one were proposed to control OB. Doping 1DPC~\cite{wang_dispersive_1997} was suggested as a compact alternative to such strategies which require increasing the size of the system. 
 
Effects of the microscopical details and possible quantum coherence of the atomic structure are not
taken into account in the general discussion of controlling OB in doped 1DPCs. On the other hand effect of atomic coherence on OB has been studied in three level atoms~\cite{walls_coherent_1980,Walls_optical_1981,Harshawardhan_controlling_1996,
anton_optical_2002,anton_optical_2003,wang_control_2009}; and it is found that 
spontaneously generated coherence (SGC) effect~\cite{javanainen_effect_1992} strongly enhances the 
nonlinear response of three level 
atoms~\cite{niu_enhancing_2006}. The effect of SGC is due to a counterintuitive role played by the vacuum modes. When two
low lying levels are separated less than the excited state line width, same vacuum modes can be emitted and re-absorbed so that quantum decay trajectories interfere to establish quantum coherence for the lower levels. This intrinsically nonlinear effect contributes to enhancement of the nonlinear response. Our objective
is to utilize this fact for efficient control of OB in doped 1DPCs. 

We consider a 1DPC with 
 $\Lambda$-type three level atoms embedded in the central layer.
Utilizing spontaneously generated coherence which enhances Kerr-type nonlinearity of the atoms in the central layer, 
we find that such a system allows for wide range control of the
contrast in optical switching and the level of power threshold. Our idea exploits first the enhancement of local intensity of light by the defect modes of the dopant atoms in the central layer of the PC, and second it exploits the additional enhancement of nonlinear
response of the atoms by SGC. Both the enhanced intensity and nonlinear response are translated into three orders of magnitude lowering of 
threshold power
to reach OB regime incontrast to the case without SGC. In addition our proposal brings flexibility to control of hysteresis loop, contrast and threshold of OB compactly 
via the atomic parameters
such as detuning and Rabi frequencies. In particular high contrast between bistable transmission states is found for a certain
set of atomic parameters. 
Significant recent technological progress to introduce dopants~\cite{braun_introducing_2006} or quantum dots~\cite{kuroda_acceleration_2008} in PCs makes our proposal promising for next generation 
photonic diodes and transistors to realize all-optical logic and memory applications. Alternatively quantum dots~\cite{dutt_stimulated_2005}, semiconductor heterostructures~\cite{wu_ultrafast_2005} or equivalent dressed state schemes~\cite{wu_control_2005} can be considered for possible implementations of our proposal.

The paper is organized as follows. In Sec.~\ref{sec:model_and_theory}, we describe our model and the method of calculation.  The multilayer 1DPC system doped with $\Lambda$-type three level atoms in the central layer is described by presenting the level scheme, SGC, linear and nonlinear susceptibilites~\cite{niu_enhancing_2006} in Sec.~\ref{sec:model}. The nonlinear transfer matrix
method~\cite{gupta_dispersive_1987,jiang_omnidirectional_2003} we employ
to calculate the probe transmission is introduced in Sec.~\ref{sec:theory}. The results are discussed in Sec.~\ref{sec:results}, where the transmission
coefficient for various atomic parameters is discussed. Possible implementation schemes of our model is discussed at the end of this section. We conclude in Sec.~\ref{sec:con}.
\section{Light Transmission through 1D PC containing a doped layer of $\Lambda$-Type three level atoms with SGC}
\label{sec:model_and_theory}
\subsection{Model System}\label{sec:model}
\begin{figure}[!t]
\begin{center}
\includegraphics[width=0.4\textwidth]{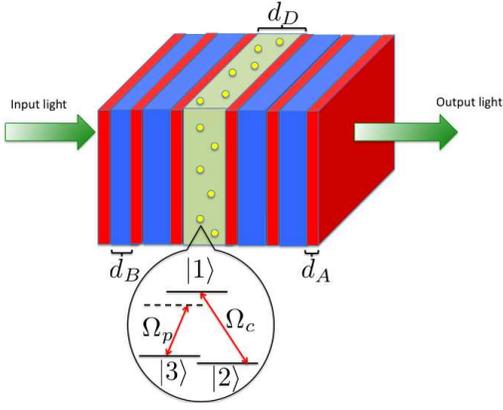}
\caption{\label{fig1} Schematic of the 1DPC we consider.  It is a symmetric multilayer stack $(AB)_2AD(AB)_2A$ where $A,B$ and $D$
are of thickness $d_A,d_B$ and $d_D$. The central layer is doped with  $\Lambda$-type three-level atoms. The coupling field drives the transitions between the levels $\vert1\rangle$ and $\vert2\rangle$with the Rabi frequency $\Omega_c$, and the probe field
drives the transitions between the levels $\vert1\rangle$ and $\vert3\rangle$ with the Rabi frequency $\Omega_p$.
We assume the coupling field is directly applied onto the doped layer $D$ while the probe field is transmitted from
the left of the 1DPC by the normal incident input field. The output is collected from right of the 1DPC structure.}
\end{center}
\end{figure}

We consider a 1DPC as depicted in Fig.~\ref{fig1} which is a symmetric multilayer stack $(AB)_2AD(AB)_2A$ where $A,B$ and $D$
are of thickness $d_A,d_B$ and $d_D$. All the layers are linear dielectrics with refractive indices $n_A,n_B$ and $n_D$. 
We take $n_A=2.22$ and $ n_D=n_B=1.41$ as in Ref.~\cite{steinberg_measurement_1993}. The parameters are related to
the midgap (central) wavelength $\lambda_{\text{pc}}= 692$ nm of the PC by $n_Ad_A=n_Bd_B=\lambda_{\text{pc}}/4$ and 
$n_Dd_D=\lambda_{pc}/2$. 
The central (defect) layer is doped with $N$ $\Lambda$-type three level atoms. A resonant coupling field  
is directly applied to the defect layer and drives the transition between the states $\vert1\rangle $ and $\vert2\rangle $
with the Rabi frequency $\Omega_c$. 
The normal incident field to the left of the 1DPC structure arrives to the defect layer and drives the transition between the states
$\vert1\rangle$ and $\vert3\rangle$ as the probe field with the Rabi frequency $\Omega_p$. 
The modified spontaneous emission rates of the atoms inside the PC are assumed to be
$2\gamma_2$ and $2\gamma_3$ from the excited state $\vert1\rangle $ to the lower states $\vert2\rangle$ and $\vert3\rangle$, respectively. 

We assume that the lower levels are closely spaced so that the two transitions to the excited state interact with the same vacuum mode and hence SGC is present. 
The first ($\chi^{(1)}$) and third-order ($\chi^{(3)}$) susceptibilities of such $\Lambda$-type three-level atoms with SGC 
atoms are given by~\cite{niu_enhancing_2006}
\begin{eqnarray}\label{eq:chi1}
\chi^{(1)} &=&-\frac{2N\vert\mu_{13}\vert^2}{\varepsilon_0\hbar}\frac{\Delta_p}{\Omega_c^2+\text{i}(\gamma_2+\gamma_3+\text{i}\Delta_p)\Delta_p},
\end{eqnarray}
\begin{eqnarray}\label{eq:chi3}
\begin{split}
\chi^{(3)}&=\frac{2N\vert\mu_{13}\vert^4}{3\varepsilon_0\hbar^3}(4i\Omega_c^4p^2\gamma_2(\gamma_2+\gamma_3)\Delta_p^2(\Omega_c^2-\Delta_p^2)\\
&+\Delta_p[\Omega_c^2+\text{i}(\gamma_2+\gamma_3+i\Delta_p)\Delta_p]\\
&\times[\Omega_c^2-\text{i}(\gamma_2+\gamma_3-i\Delta_p)\Delta_p]\\
&\times\{2\Omega_c^4\gamma_3-\text{i}\Omega_c^2\gamma_2(\gamma_2+\gamma_3)\Delta_p\\
&+\Delta_p^2[3\Omega_c^2\gamma_2+\gamma_2\gamma_3^2+\gamma_2^3\\
&+2\gamma_3(\Omega_c^2+\gamma_2^2)]\})/\beta,
\end{split}
\end{eqnarray}
with
\begin{eqnarray}
\begin{split}
\beta&=\gamma_3[\Omega_c^3-\text{i}\Omega_c(\gamma_2+\gamma_3-i\Delta_p)\Delta_p]^2\\
&\times[\Omega_c^2+\text{i}(\gamma_2+\gamma_3+i\Delta_p)\Delta_p]^3.
\end{split}
\end{eqnarray}
where $\triangle_p=\omega_{13}-\omega_p$ is the detuning of the probe field frequency $\omega_p$ from the atomic transition 
resonance at $\omega_{13}$ between the levels $\vert 1\rangle$ and $\vert 3\rangle$. The Rabi frequencies 
$\Omega_c $ and $ \Omega_p $ are constrained as $ \Omega_{c(p)}=\Omega_{c(p)}^0\sin\theta=\Omega_{c(p)}^0\sqrt{1-p^2} $ where 
$p=\cos\theta=\vec{\mu}_{12}.\vec{\mu}_{13}/\vert\vec{\mu}_{12}.\vec{\mu}_{13}\vert$ is defined as the SGC parameter for 
the two dipole moments $  \vec{\mu}_{12}$ and $ \vec{\mu}_{13} $ making an angle $ \theta $ with each other. We distinguish the
Rabi frequency for the transverse aligned dipoles ($\theta=90^\circ$) as $\Omega_{c(p)}^0$. The constraint arises by the requirement that the probe and the coupling fields do not interact with each other's transitions so that one must be perpendicular to the dipole moment coupled to the other. In general Rabi frequencies are complex numbers but we shall take them real valued here for simplicity. Mathematically
the relative phase between the probe and the coupling fields can be studied by considering a complex valued SGC parameter and may lead to multistability beyond OB. We shall not include this case to our considerations. 
\subsection{Nonlinear transfer matrix method}\label{sec:theory}
We employ the standard characteristic (transfer) matrix method to calculate the transmission coefficient for our 1DPC system~\cite{gupta_dispersive_1987,agarwal_effect_1987,gupta_optical_1988}. 
The transfer matrix $M$ for the multilayer structure in Fig.~\ref{fig1} is written by
\begin{eqnarray}
M&=(M_AM_B)^2M_AM_D(M_AM_B)^2M_A,
\end{eqnarray}
where $M_j$ with $j=A,B,D$ are the transfer matrices for the corresponding layers.
The transmission coefficient $T$ is the ratio of the transmitted field intensity to the incident field intensity and it is related to the elements $m_{ij}$ with $i,j=1,2$ of the transfer matrix $M$ by
\begin{eqnarray}
T&=\left\vert\frac{2n_0}{(m_{11}+m_{12}n_0)+(m_{21}+m_{22}n_0)}\right\vert,
\end{eqnarray}
where $n_0$ is the refractive index of the air. 
We assume the input field is incident from the air and the output field is transmitted into the air. 

When a TE-polarized normal incident pulse is considered, transfer matrix for the defect layer $D$, doped with nonlinear atoms, 
is given by~\cite{gupta_dispersive_1987,gupta_optical_1989}   
\begin{eqnarray}
M_D=
\left(
\begin{array}{cc}
m_{11}^{(d)}&m_{12}^{(d)}\\
m_{21}^{(d)}&m_{22}^{(d)},
\end{array}
\right)
\end{eqnarray}
where
\begin{eqnarray}
m_{11}^{(d)}&=\frac{1}{k_++k_-}\left( k_-\text{e}^{-\text{i}k_{+}d_D}+k_+\text{e}^{\text{i}k_{-}d_D} \right),\\
m_{12}^{(d)}&=\frac{k_0}{k_{+}+k_{-}}\left(\text{e}^{-\text{i}k_{+}d_D}-\text{e}^{\text{i}k_{-}d_D}\right),\\
m_{21}^{(d)}&=\frac{k_-k_+}{k_0(k_{+}+k_{-})}\left(\text{e}^{-\text{i}k_{+}d_D}-\text{e}^{\text{i}k_{-}d_D}\right),\\
m_{22}^{(d)}&=\frac{1}{k_{+}+k_{-}}\left(k_+\text{e}^{-\text{i}k_{+}d_D}+k_-\text{e}^{\text{i}k_{-}d_D}\right).
\end{eqnarray}
Here the propagation constants of the forward and backward propagating probe field inside the doped layer $D$ are
denoted by $k_{+}$ and $k_{-}$ and they depend on the field amplitudes due to the nonlinear dopant atoms by
\begin{eqnarray}
k_{\pm}=k_0n_l(1+U_{\pm}+2U_{\mp})^\frac{1}{2},
\end{eqnarray}
where $k_0$ is the wave vector in vacuum, and
\begin{eqnarray}\label{eq:Upm}
U_{\pm}&=\chi^{(3)}\vert A_{\pm}\vert^2,
\end{eqnarray}
where $ A_{j\pm} $ and $ A_{j\pm} $ are the amplitudes of the forward and backward propagating probe fields and $n_l$ is the linear refractive index of the layer $D$ including the dopant atoms. 
Dielectric permittivity of defect layer $ (\epsilon) $ can be written in the form of $\epsilon=\epsilon_l+\chi^{(3)} \vert E(z)\vert^2$ with $ \epsilon_l $ is the linear dielectric permittivity. For our system (a linear defect layer doped with $\Lambda$-type three level atoms)  linear dielectric permittivity can be calculated by 
\begin{eqnarray}
\epsilon_l=\epsilon_D+\chi^{(1)}.
\end{eqnarray}
Here $ \epsilon_D=n_D^2 $ is dielectric permittivity of linear defect layer. 

Nonlinear character of the dopant atoms makes the wave vectors $ k_{\pm} $ depend on the forward and backward propagating probe field
intensities $U_{\pm}$ inside the central layer. In order to construct $ M_D $ we need to determine $U_\pm$ by solving 
a set of coupled nonlinear equations,
\begin{eqnarray}
U_\pm=\left\vert\frac{p_+(m_{11}^{(r)}+m_{12}^{(r)})\pm(m_{21}^{(r)}+m_{22}^{(r)})}
{p_-+p_+}\right\vert^2U_f,
\end{eqnarray}
with 
\begin{eqnarray}
p_\pm=n_l\sqrt{1+U_\mp+2U_\pm}
\end{eqnarray}
for a given transmitted intensity $U_f$ (scaled by $\chi^{(3)}$) by the fixed point iteration method~\cite{gupta_dispersive_1987}. Here we denote
the elements of the transmission matrix for the multilayer stack to the right of the central layer, $M_r=(M_AM_B)^2M_A$
as $m_{ij}^{(r)}$. These equations reflect the relation between the tangential field components at the right boundary of the
central layer and at the output surface of the 1DPC. 

For the layers $A$ and $B$, the transfer matrices $M_A$ and $M_B$ are given by~\cite{jiang_omnidirectional_2003}, 
\begin{eqnarray}
M_j&=\left( \begin{array}{cc}
\cos(k_jd_j) & -\text{i}\frac{\sqrt{\mu_j}}{\sqrt{\epsilon_j}}\sin(k_jd_j)\\
-\text{i}\frac{\sqrt{\epsilon_j}}{\sqrt{\mu_j}}\sin(k_jd_j) & \cos(k_jd_j)\end{array} \right),
\end{eqnarray}
where $ k_j=\sqrt{\epsilon_j\mu_j}\omega_p/c$ with $j=A,B$. Using the relation between the incident intensity $U_{\text{i}}$ (scaled by $\chi^{(3)}$) and the 
transmitted intensity $U_f$ by $U_{i}=U_f/T$ we determine the dependence of the transmission to the $U_i$.

As the relations among the intensities $U_\pm,U_f$ and $U_{i}$ are scaled by the nonlinear susceptibility $\chi^{(3)}$,
the explicit dependence of the transfer matrix to the atomic parameters are only due to the linear index $n_l$. To control OB
efficiently both the linear and nonlinear susceptibilities hence should be carefully considered together. In the next section we 
numerically evaluate the transmission coefficient to reveal the effects of atomic parameters and SGC 
on the OB in the light of the linear and nonlinear responses of the dopant atoms.
\section{Results and Discussion}\label{sec:results}
Our purpose is to examine the influence of atomic parameters, specifically the probe detuning, coupling field Rabi frequency,
and the SGC parameter, on the OB. For that aim it is convenient for us to write the expressions of the linear and nonlinear susceptibilities 
in forms that explicitly depends on these parameters {\it per se}. We take equal decay rates from the excited state to the closely spaced doublet, $\gamma_2=\gamma_3:=\gamma$, for simplicity and 
divide the numerators and denominators of Eq.~(\ref{eq:chi1}) and 
Eq.~(\ref{eq:chi3}) by $\gamma$ and $\gamma^{13}$, respectively.

\begin{figure}[!t]
\includegraphics[width=0.4\textwidth]{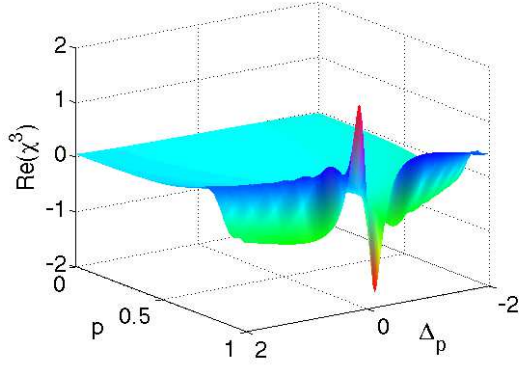}
\caption{\label{fig2} (Color online) Dependence of the real part of the nonlinear susceptibility $\text{Re}(\chi^{3})$ (in arbitrary units (a.~u.)) on the SGC parameter $p$ and the probe detuning $\Delta_p$ for $\Omega_c^0=4$. Significant enhancement of the nonlinear response of the system can be obtained at the optimal value of $p\sim 1$.($\Delta_p$ and $\Omega_c^0$ are dimensionless and scaled by the spontaneous decay rate $\gamma$.)}
\end{figure}

We use $\gamma$ to scale quantities in frequency units, especially $\Omega_c, \Delta_p$, and to make them dimensionless, so that we can write
$\chi^{(1)}=s^{(1)}\tilde{\chi}^{(1)}$ and $\chi^{(3)}=s^{(3)}\tilde{\chi}^{(3)}$ where the factors $s^{(1)},s^{(3)}$  
\begin{eqnarray}
s^{(1)}=\frac{2N\vert{\mu_{13}\vert^2}}{\varepsilon_0\hbar\gamma},
\quad
s^{(3)}=\frac{2N\vert\mu_{13}\vert^4}{3\varepsilon_0\hbar^3\gamma^3},
\end{eqnarray}
depending on the atomic constants are separated from the dimensionless factors $\tilde{\chi}^{(1)}, \tilde{\chi}^{(3)}$ 
\begin{eqnarray}
\tilde{\chi}^{(1)}&=&-\frac{\Delta_p}{\Omega_c^2+\text{i}(2+\text{i}\Delta_p)\Delta_p},\\
\tilde{\chi}^{(3)}&=&(8i\Omega_c^4p^2\Delta_p^2(\Omega_c^2-\Delta_p^2)\\
&+&\Delta_p[\Omega_c^2+\text{i}(2+i\Delta_p)\Delta_p]\\
&\times&[\Omega_c^2-\text{i}(2-i\Delta_p)\Delta_p]
\times\{2\Omega_c^4-\text{i}2\Omega_c^2\Delta_p\\
&+&\Delta_p^2[3\Omega_c^2+2+2(\Omega_c^2+1)]\})/\tilde{\beta},
\end{eqnarray}
with
\begin{eqnarray}
\tilde{\beta}&= &[\Omega_c^3-\text{i}\Omega_c(2-i\Delta_p)\Delta_p]^2\\
&\times&[\Omega_c^2+\text{i}(2+i\Delta_p)\Delta_p]^3,
\end{eqnarray}
depending on the
control parameters.
Using the relation between the dipole moment and the decay rate
$\gamma=\vert{\mu_{13}\vert^2}\omega_{13}^3/3\pi\varepsilon_0\hbar c^3$, it can be verified immediately that $s^{(1)}=6\pi Nc^3/\omega_{13}^3$ is dimensionless while $s^{(3)}=s^{(1)}\pi\varepsilon_0c^3/\omega_{13}^3\hbar\gamma$ has units of inverse electric field squared. 

\begin{figure*}[!hbt]
\subfloat[][]{
\includegraphics[width=0.4\textwidth]{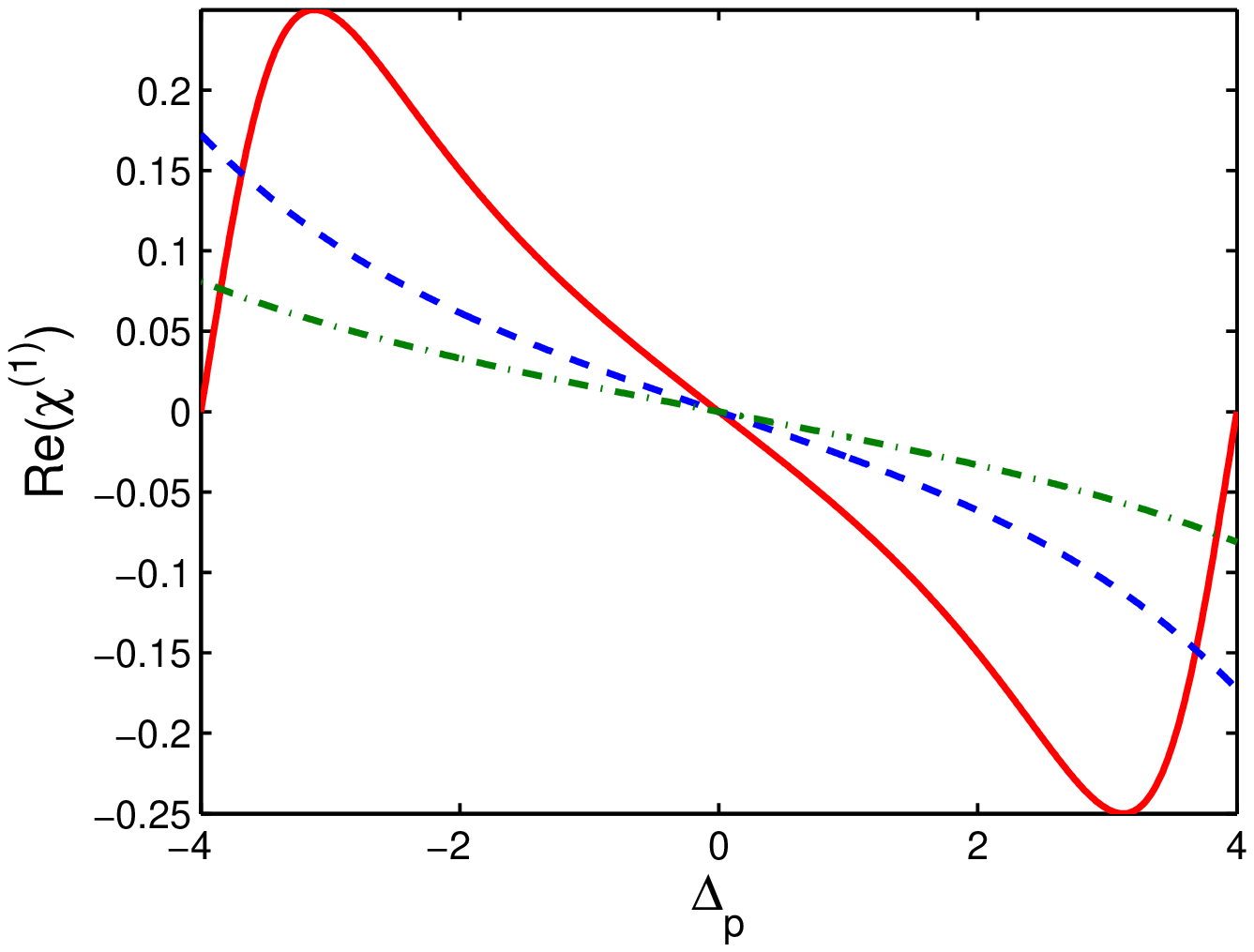}
\label{fig3a}}
\subfloat[][]{
\includegraphics[width=0.4\textwidth]{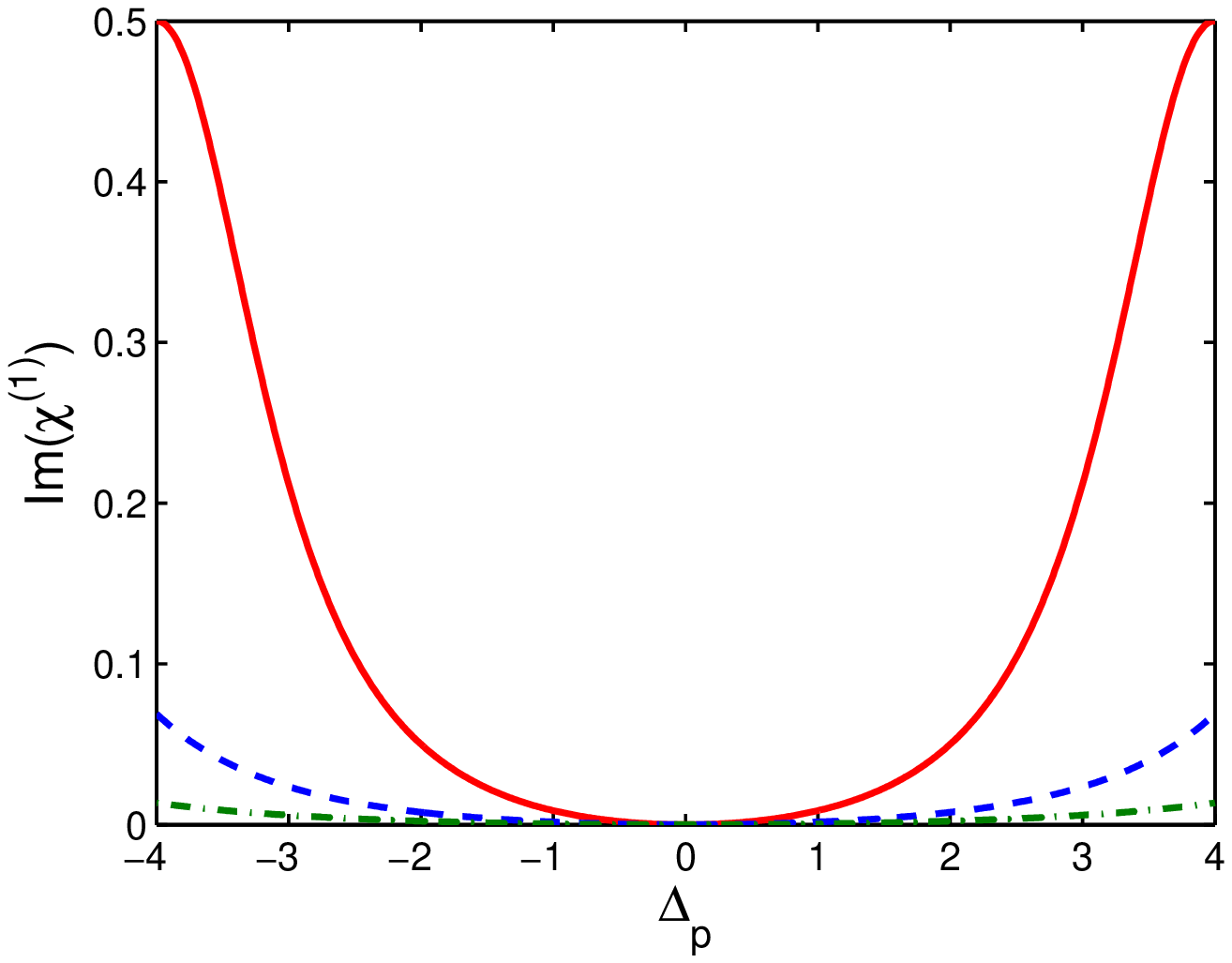}
\label{fig3b}}
\qquad
\subfloat[][]{
\includegraphics[width=0.4\textwidth]{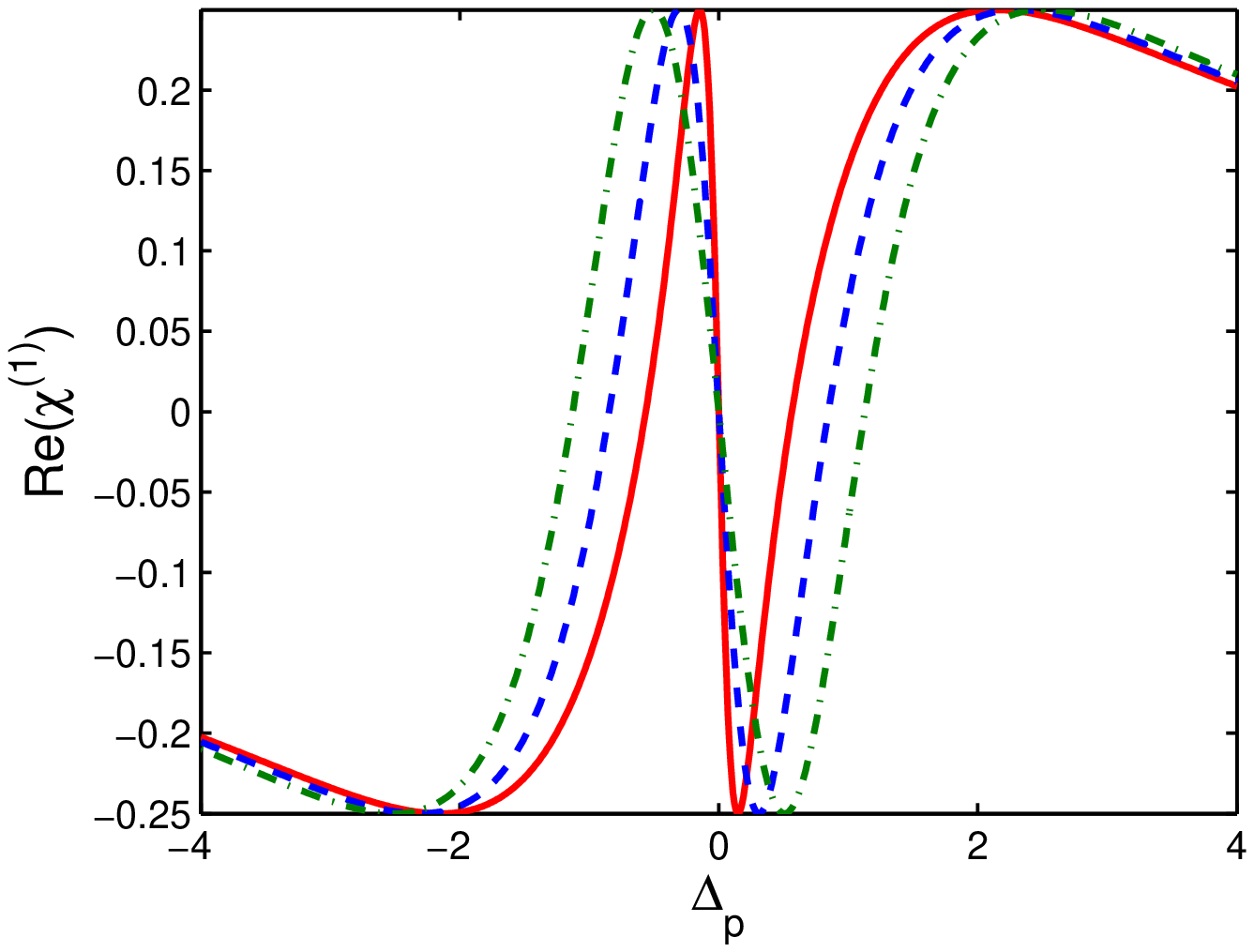}
\label{fig3c}}
\subfloat[][]{
\includegraphics[width=0.4\textwidth]{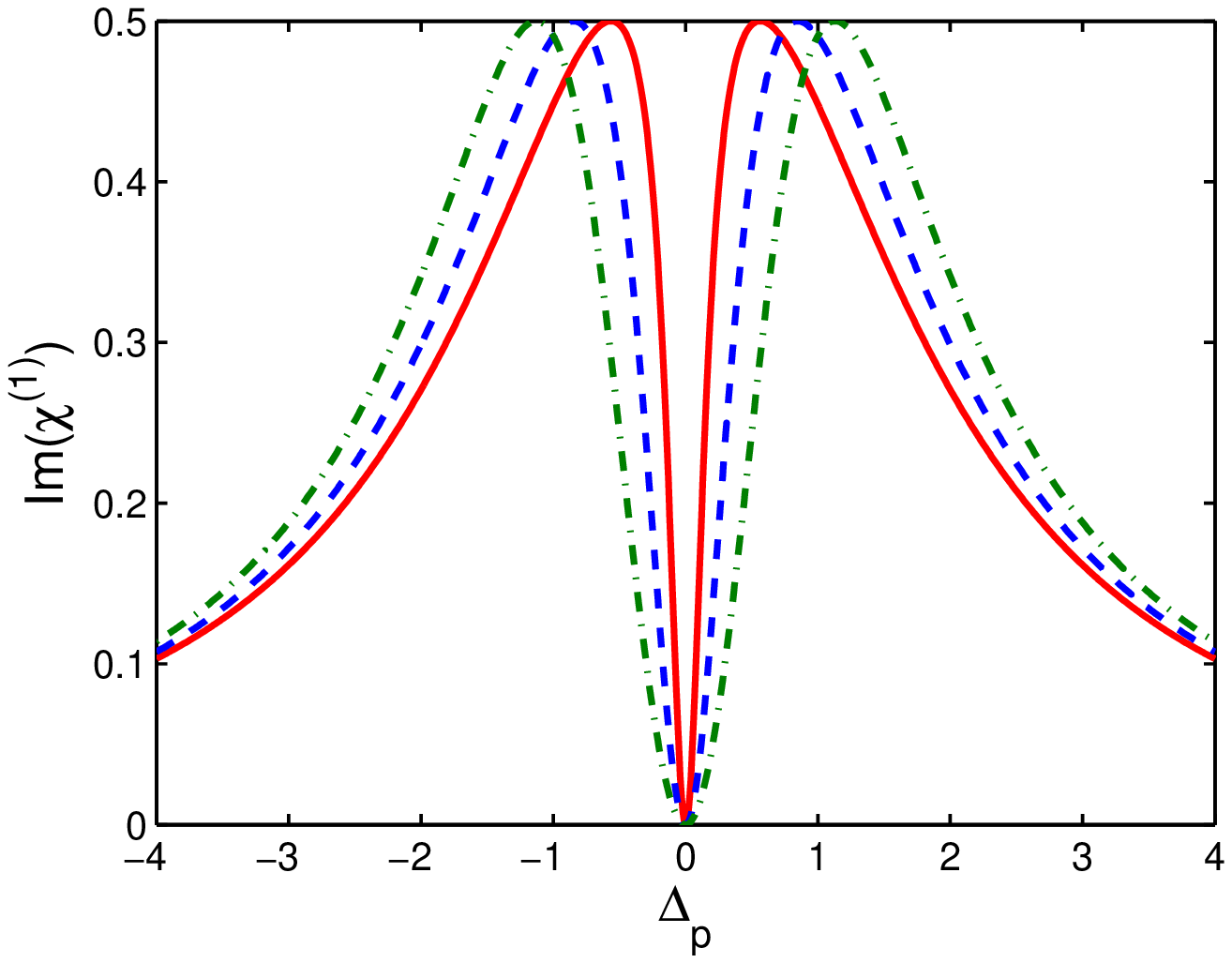}
\label{fig3d}}
\caption{\label{fig3} (Color online) 
(a) Real $\text{Re}\chi^{(1)}$ and (b) imaginary $\text{Im}\chi^{(1)}$ parts of the linear susceptibility   
as a function of probe detuning $\Delta_p$ when SGC is absent ($p=0$). (c) Real $\text{Re}\chi^{(1)}$ and (d) imaginary $\text{Im}\chi^{(1)}$ parts of the linear susceptibility as a function of probe detuning $\Delta_p$ when SGC is present ($p=0.99$). Curves with red solid line, blue dashed line, and green dash-dotted line correspond to the 
$\Omega_c^0=4,\Omega_c^0=6$, and $\Omega_c^0=8$, respectively. ($\Delta_p$ and $\Omega_c^0$ are dimensionless and scaled by the spontaneous decay rate $\gamma$.)
} 
\end{figure*}

\begin{figure*}[!t]
\subfloat[][]{
\includegraphics[width=0.4\textwidth]{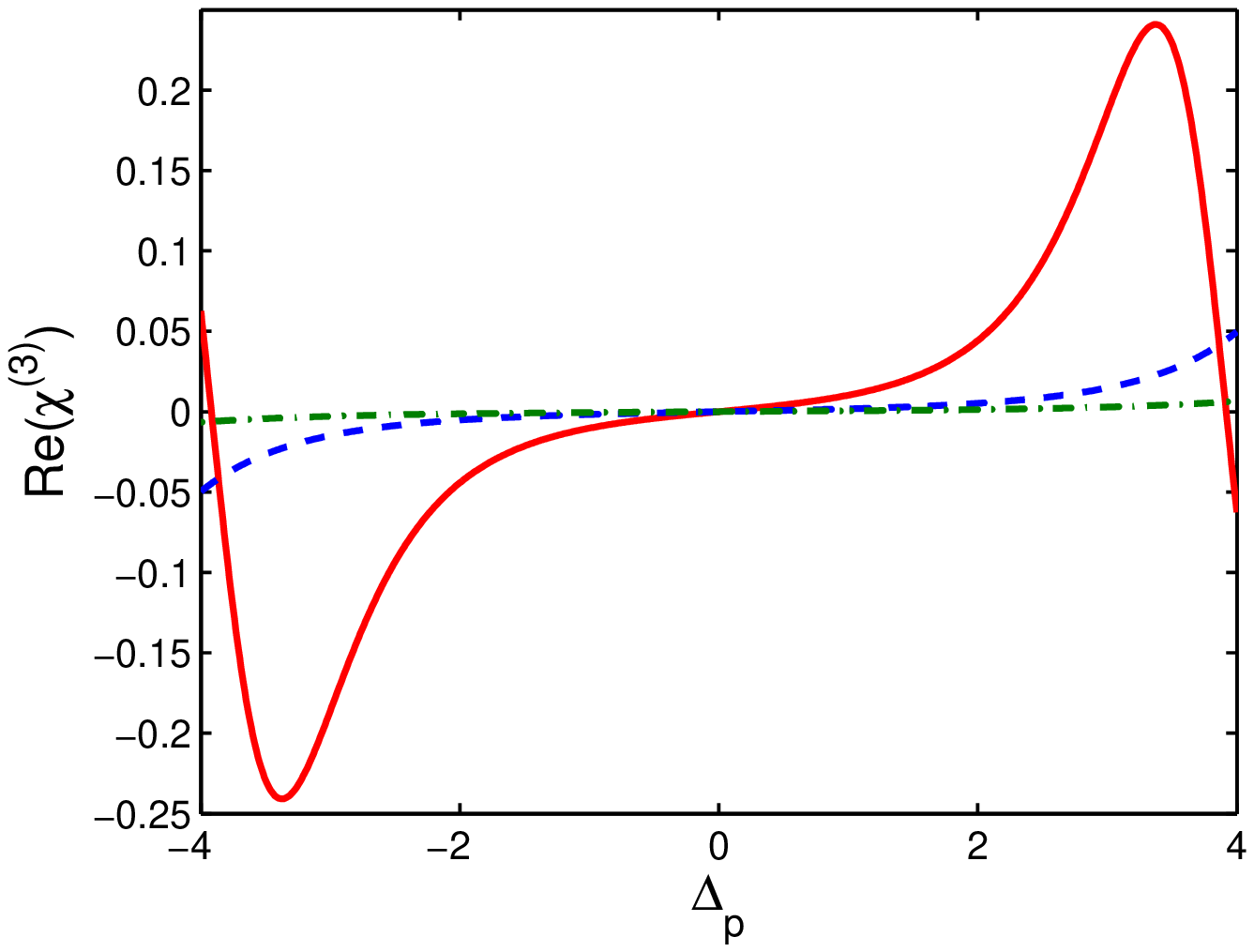}
\label{fig4a}}
\subfloat[][]{
\includegraphics[width=0.4\textwidth]{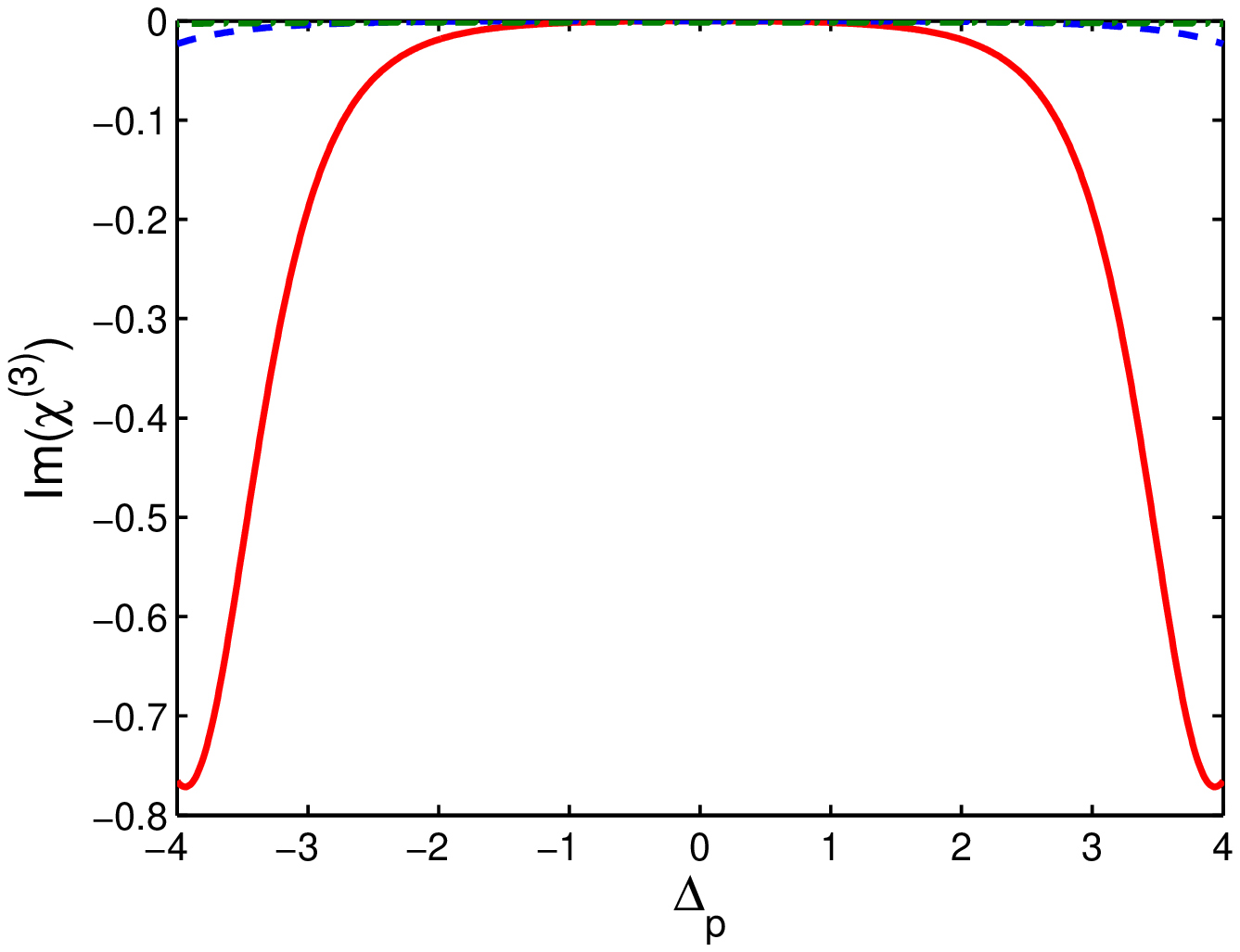}
\label{fig4b}}
\qquad
\subfloat[][]{
\includegraphics[width=0.4\textwidth]{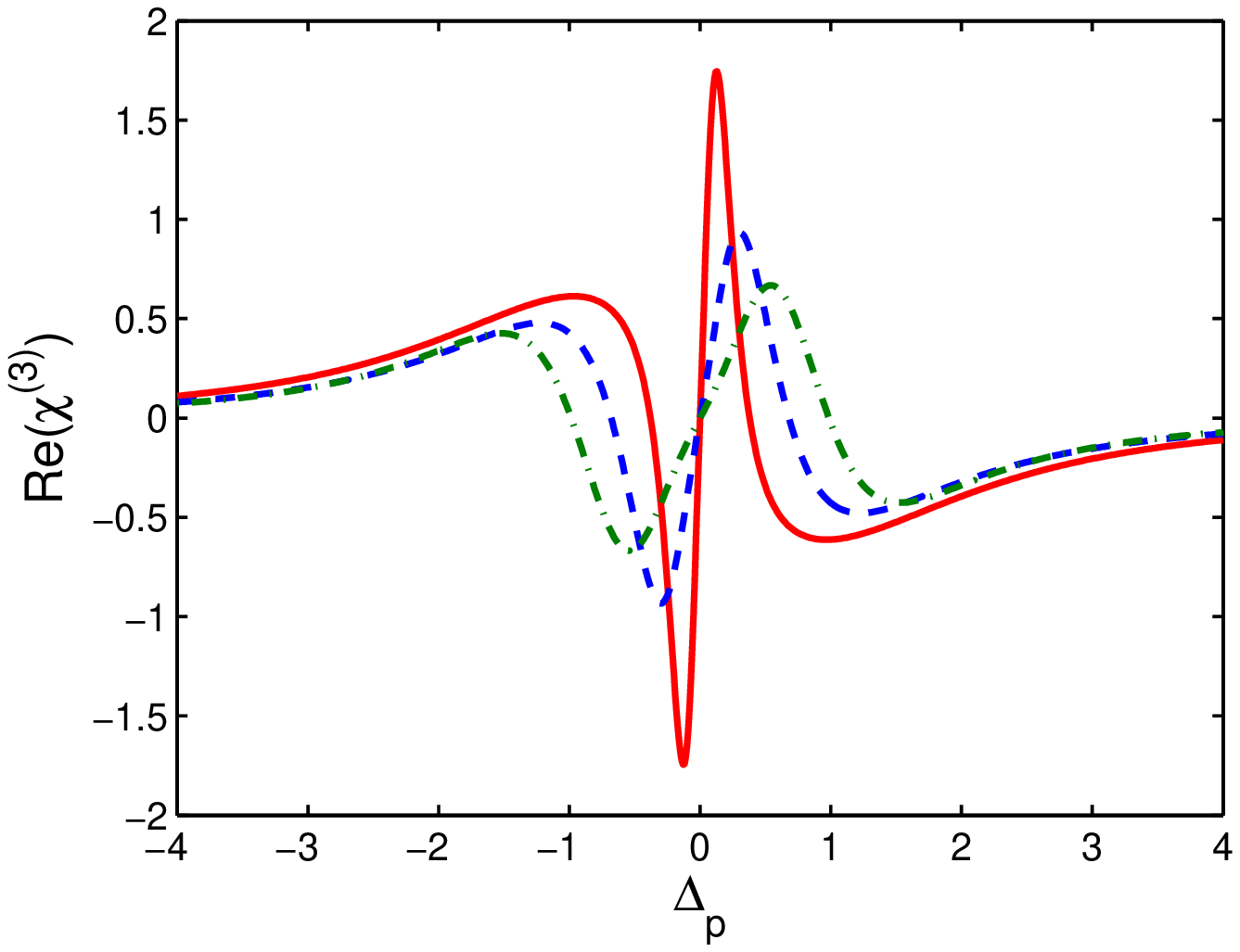}
\label{fig4c}}
\subfloat[][]{
\includegraphics[width=0.4\textwidth]{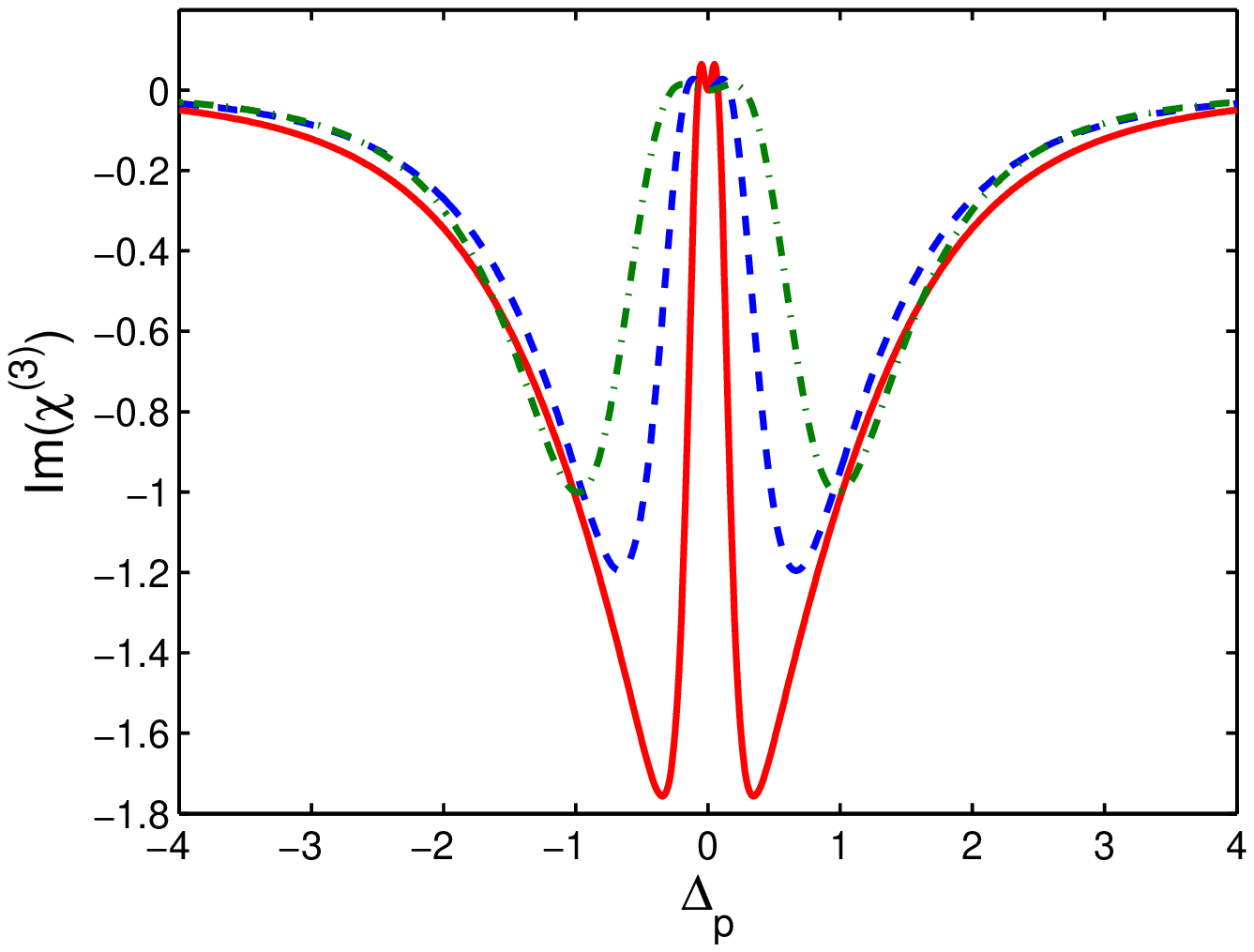}
\label{fig4d}}
\caption{\label{fig4} (Color online)
(a) Real $\text{Re}\chi^{(3)}$ and (b) imaginary $\text{Im}\chi^{(3)}$ parts of the nonlinear susceptibility (dimensionless) 
as a function of probe detuning $\Delta_p$ when SGC is absent ($p=0$). (c) Real $\text{Re}\chi^{(3)}$ and (d) imaginary $\text{Im}\chi^{(3)}$ parts of the nonlinear susceptibility  (dimensionless) as a function of probe detuning $\Delta_p$ when SGC is present ($p=0.99$). Curves with red solid line, blue dashed line, and green dash-dotted line correspond to the 
$\Omega_c^0=4,\Omega_c^0=6$, and $\Omega_c^0=8$, respectively. ($\Delta_p$ and $\Omega_c^0$ are dimensionless and scaled by the spontaneous decay rate $\gamma$.)
} 
\end{figure*}

We apply the nonlinear transfer matrix method to determine the transmission of an incident probe light with carrier frequency $\omega_p=2.5\times10^{15}$ Hz near to the midgap frequency $\omega_{pc}=2\pi c/\lambda_{pc}=2.72\times10^{15}$ Hz of the 1DPC. Doped 1DPC supports a linear defect mode around the midgap
frequency and allows for a linear transmission window within the band gap. In order to make nonlinear response relevant for the probe
transmission we consider the following scheme. The atomic resonance is assumed to be slightly detuned 
from the probe and thus lies within the photonic band gap but
just below the transmission window opened by the linear defect mode. The atomic
spontaneous decay rate would then be modified~\cite{bravo-abad_enhanced_2007} but we do not need its
actual value as we use it for scaling factor. Typical range of values for the control parameters we use in our
simulations relative to $\gamma$ are as follows. Detuning of the incident light from the atomic probe transition is assumed to be in the range from
$\Delta_p=0.05$ to $\Delta_p=0.15$. Control field Rabi frequency $\Omega_c^0$ is considered to be in the range $\Omega_c^0\sim 1-10$. We also assume background material as air surrounding the non-magnetic dielectric layers so that $n_0=1,\mu_{A,B,D}=1$. 

While we use $\chi^{(3)}$ when we investigate the transmission as a function of the incident field intensity $I_i$; it only indirectly appears in the treatment
of nonlinear transfer matrix method, as a scaling factor within $U_\pm$ in Eq.~(\ref{eq:Upm}). The linear susceptibility however is directly and explicitly used in the calculations. Due to these two distinct stages of using $\chi^{(1)}$ and $\chi^{(3)}$,
we drop the factor $s^{(3)}$ in $\chi^{(3)}$ by assuming the
intensity is measured in arbitrary units (a.u.). 
In the following we shall drop the tilde notation and use 
$\chi^{(3)}$ in place  of $\tilde{\chi}^{(3)}$.
We further assume that $s^{(1)}=1$, which can be satisfied by taking a proper $N$. In our case $\omega_{13}\sim 10^{15}$ Hz so that $N\lesssim 10^{20}$ m$^{-3}$. For the same parameters $s^{(3)}\sim 5.4\times 10^{-8}$ m$^2$/V$^2$. Intensity can be converted from a.u. to physical units
of mW$/$cm$^2$ for these atomic variables by a multiplication factor of $50$. Despite this theoretical association of a.u. with the physical units here, we follow the common conventional wisdom to give intensity in a.u. due to its specific measurement dependence.

We first examine the behavior of $\chi^{(3)}$ to determine the optimal set of parameters to get significant nonlinear response out of the doped 1DPC system. Enhancement of the nonlinearity by the SGC can be seen in Fig.~\ref{fig2} where the largest value of $\chi^{(3)}$ is found at $\Delta_p\sim 0.05$ for $p\sim 1$. Such a large nonlinear response can be exploited for OB if the absorption is weak in the corresponding frequency window.

\begin{figure}[!t]
\includegraphics[width=0.4\textwidth]{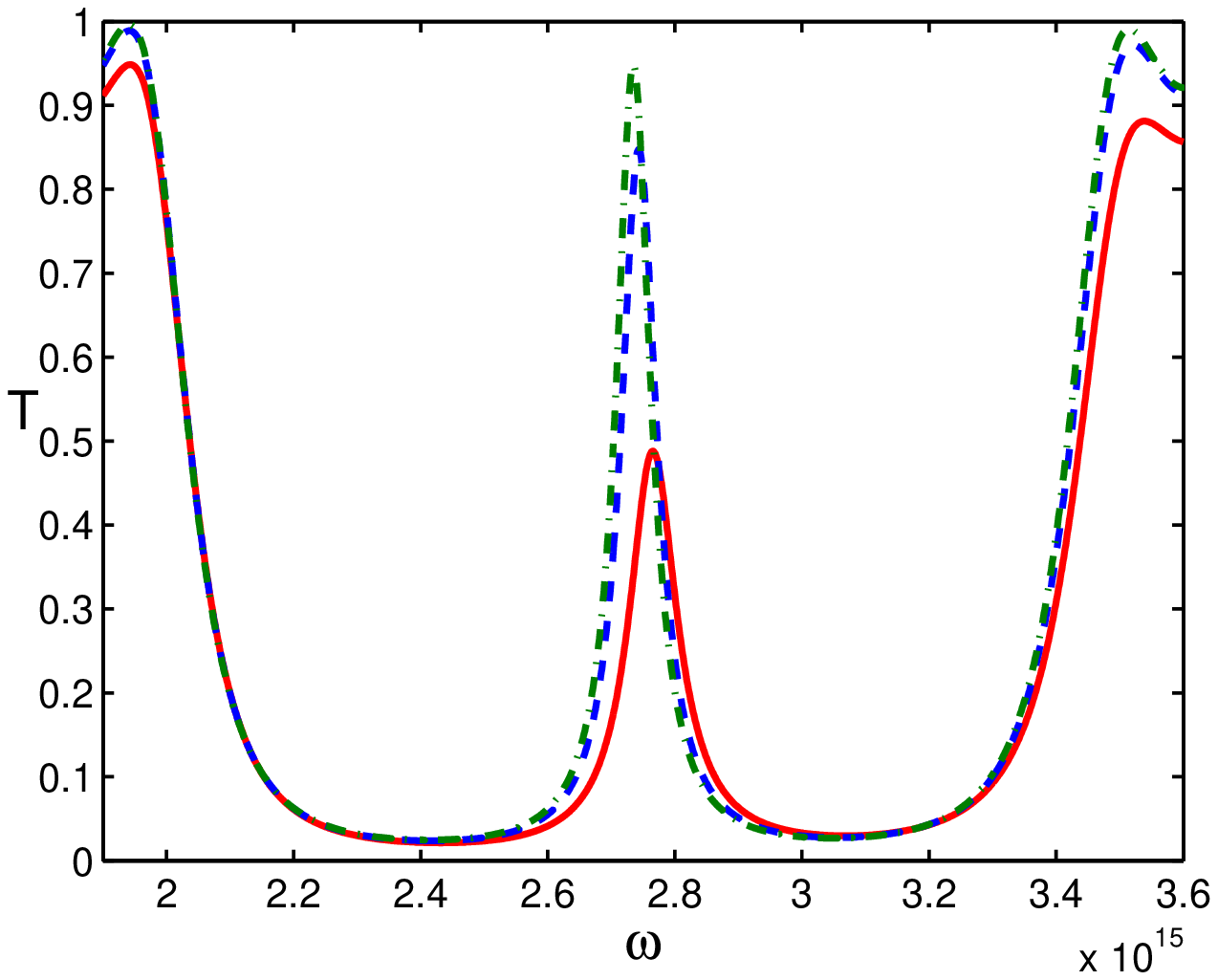}
\caption{\label{fig5} (Color online)
Linear transmission coefficient $T$ of a 1DPC including a central defect layer doped with $\Lambda$-type three level atoms with SGC as a function of the incident field frequency $\omega$ (in Hz) for $ \Omega_c^0 = 4$ (blue 
dashed curve), $\Omega_c^0=6$ (green dash-dotted curve) and $\Omega_c^0=8$ (red solid curve). Other parameters are $\Delta_p=0.05$ and $p=0.99$. ($\Delta_p$ and $\Omega_c^0$ are dimensionless and scaled by the spontaneous decay rate $\gamma$.)} 
\end{figure}

We subsequently explore the linear susceptibility and absorption properties of the system in Fig.~\ref{fig3} and in Fig.~\ref{fig4} where we
compare the cases of lack of SGC ($p=0$) and presence of optimal ($p=0.99$) SGC. These figures also explore the effect of the coupling laser Rabi frequency. The curves with the red solid line, blue dashed line and the green dash-dotted line correspond to $\Omega_c^0=4,\Omega_c^0=6$, and $\Omega_c^0=8$, respectively.

Real part of the linear susceptibility exhibits steep negative dispersion near the probe resonance if SGC is present, as can be seen by comparing Fig.~\ref{fig3a} with Fig.~\ref{fig3c}. This is consistent with the superluminal light propagation and the Hartman effect which are enhanced by the SGC according to recent investigations of 1DPC doped with $\Lambda$-type three level atoms~\cite{sahrai_hartman_2013}. Barrier length independence of the tunneling time~\cite{maccoll_note_1932}, known as the Hartman effect~\cite{hartman_tunneling_1962}, in the doping region can be beneficial for OB based fast optical switching applications for practical implementation of our model system. When we investigate the imaginary part of the linear susceptibility in Fig.~\ref{fig3b} and in Fig.~\ref{fig3d}, we see that the transmission window becomes more narrow in the presence of SGC. Nevertheless, strong nonlinearity regime, in particular around the optimal detuning of $\Delta_p\sim 0.05$, remains in the narrow transmission window. The negative dispersion gets steeper and the width of the transmission window gets smaller with the decrease of the $\Omega_c^0$.

These conclusions hold true when we consider the nonlinear dispersion and absorption curves in Fig.~\ref{fig4}. Real part of the nonlinear susceptibility plotted in Fig.~\ref{fig4a} and in Fig.~\ref{fig4c} show that the steep changes around the probe 
resonance emerge in the presence of SGC. In contrast to anomalous linear dispersion, nonlinear one is normal with positive slope. More crucial observation for the ease of OB operation is the significant enhancement of the magnitude of
$\chi^{(3)}$ with the SGC. About an order of magnitude increase 
is obtained at $p=0.99$. The nonlinear transmission window is 
narrowed as well. The detuning required for the strong nonlinear
response is within the narrow window of transmission. An additional information to similar Fig.~\ref{fig2} available here is the effect of $\Omega_c^0$. The nonlinear dispersion gets steeper and the nonlinear transmission window gets narrower with the decrease of $\Omega_c^0$. Accordingly these observations suggest that we can make a further optimal choice for large nonlinear response in a transmission window by taking $\Omega_c^0=4$ at $\Delta_p=0.05$. 

\begin{figure}[!t]
\subfloat[][]{
\includegraphics[width=0.4\textwidth]{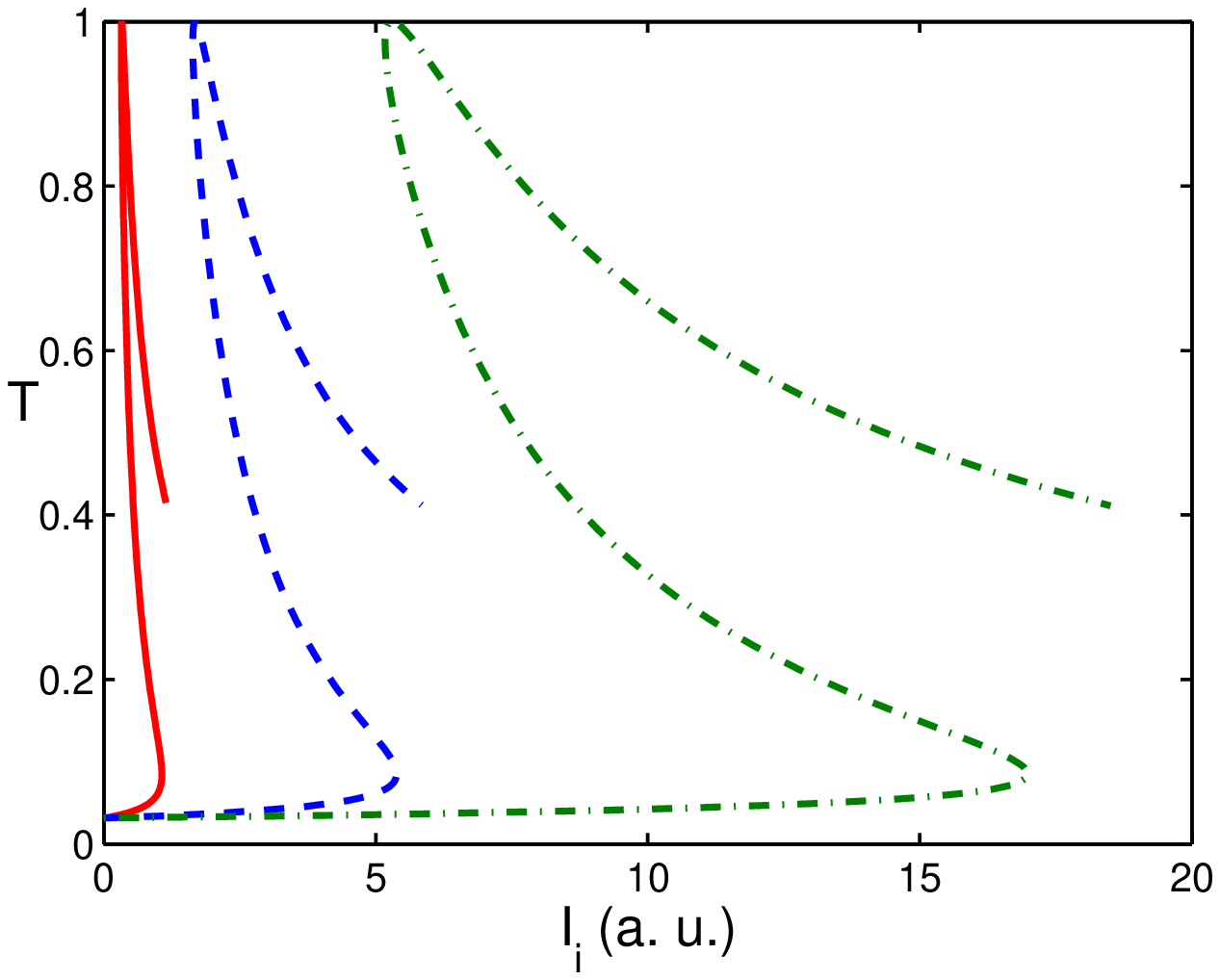}
\label{fig6a}}
\qquad
\subfloat[][]{
\includegraphics[width=0.4\textwidth]{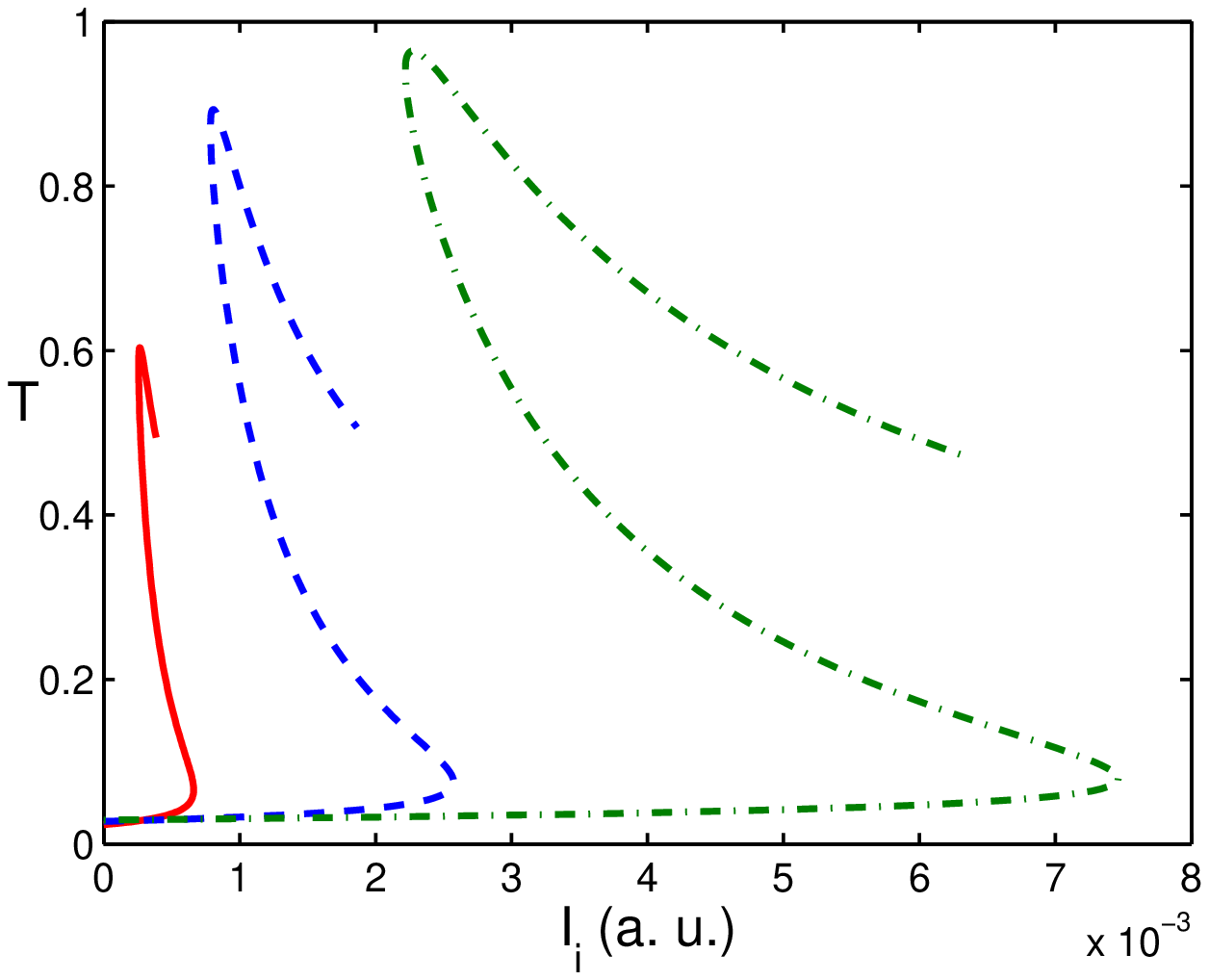}
\label{fig6b}}
\caption{\label{fig6} (Color online) 
Transmission coefficient $T$ of a 1DPC including a central defect layer doped with $\Lambda$-type three level 
atoms with SGC as a function of the incident light intensity $I_i$ (in arbitrary units (a.u.)) at probe detuning $\Delta_p=0.05$ 
when (a) SGC is absent ($p=0$) and when (b) SGC is optimal ($p=0.99$).
Different curves with red solid line, blue dashed line, and green 
dash-dotted line correspond to the 
$\Omega_c^0=4,\Omega_c^0=6$, and $\Omega_c^0=8$, respectively. ($\Delta_p$ and $\Omega_c^0$ are dimensionless and scaled by the spontaneous decay rate $\gamma$.)
} 
\end{figure}

If we calculate the linear transmission spectrum (by taking $\chi^{(3)}=0$) we can notice the emergence of the defect mode within
the photonic band gap. The result, for different values of Rabi frequencies of the control field, is plotted in Fig.~\ref{fig5}. The transmission resonances become narrower for smaller $\Omega_c^0$.
It is expected that it is easier to achieve OB with sharp defect modes ~\cite{agarwal_exact_1986}.  
Typical solution to achieve such narrow Airy resonances in Fabry-Perot
type systems is to enlarge the number of coupled resonators or photonic crystal coatings~\cite{gupta_dispersive_1987}. Our atomic control parameter $\Omega_c^0$
could potentially be a more compact solution than increasing the layers in the 1DPC. On the other hand
the effect of $\Omega_c^0$ on OB is not trivial. In addition to its influence on the width of the defect mode,
an accompanying frequency shift of the defect mode from the
probe frequency can also be seen in the Fig.~\ref{fig5}. It is known that
high frequency shift of the linear defect mode yields higher OB
threshold~\cite{wang_dispersive_1997}. The width of the defect mode becomes sharper while its frequency shifts higher away from the probe resonance with the decrease of $\Omega_c^0$. In our case therefore beneficial and harmful effects of $\Omega_c^0$ on OB compete. We need to investigate the nonlinear transmission carefully to assess the net effect of $\Omega_c^0$ explicitly.

Including the full linear and nonlinear response, the transmission coefficient $T$ is calculated and plotted as a function of the input field intensity $I_i$ (in arbitrary units (a.u.)) for different Rabi frequencies $\Omega_c^0$ of the coupling field in Fig.~\ref{fig6}. We see that by decreasing the Rabi frequency of coupling field $\Omega_c^0$ the threshold of the OB is decreased. Comparison of Fig.~\ref{fig6a} and Fig.~\ref{fig6b}
shows that SGC significantly lowers the OB threshold while keeping the hysteresis loop size the same. Three orders of magnitude decrease in the threshold intensity can be achieved. Maximum transmission or the contrast between the lowest and highest points in the hysteresis loop is decreased with the increasing SGC. This effect is stronger fow smaller $\Omega_c^0$. An optimal choice of the Rabi frequency of the coupling field and the SGC parameter can be made depending on the requirements of particular applications.

\begin{figure}[!t]
\begin{center}
\includegraphics[width=0.4\textwidth]{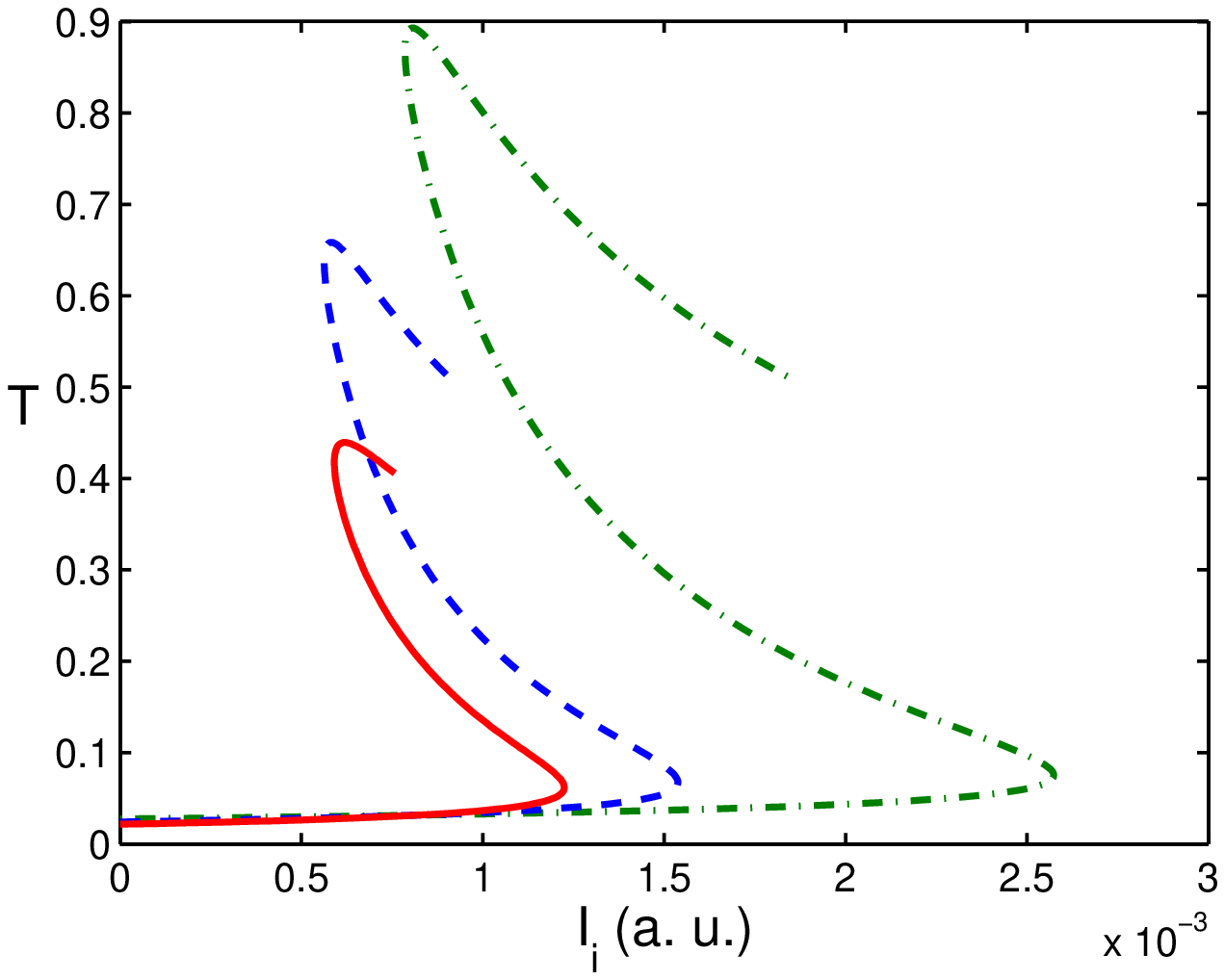}
\caption{\label{fig7}  (Color online)
Transmission coefficient $T$ of a 1DPC including a central defect layer doped with $\Lambda$-type three level 
atoms with SGC as a function of the incident light intensity $I_i$ (in arbitrary units (a.u.)) at coupling field Rabi frequency $\Omega_c^0=6$ 
and when SGC is optimal ($p=0.99$).
Different curves with blue dashed line, green dash-dotted line and red solid line correspond to the 
$\Delta_p=0.05, \Delta_p=0.1$, and $\Delta_p=0.15$, respectively. ($\Delta_p$ and $\Omega_c^0$ are dimensionless and scaled by the spontaneous decay rate $\gamma$.)} 
\end{center}
\end{figure}

Nonlinear transfer matrix method produces the relation between $T$ and scaled dimensionless incident field intensity $U_i$. We used the expression $I_i=c\varepsilon_0 U_i/2\mathrm{Re}(\chi^{(3)})$ for the 
incident field intensity to plot the Figs.~\ref{fig6}-\ref{fig7}. Dependence of $\chi^{(1)}$ on the SGC parameter $p$ and $\Omega_c^{0}$ via $\Omega_c$ results in differences in the values of $U_i$ at the OB threshold. These differences are further changed by $\chi^{(3)}$ when $U_i$ is translated to the physical intensity $I_i$. A comparison of $\chi^{(1)}$, $\chi^{(3)}$ and
$U_i$ can be made by examination of Table~\ref{table1} which confirms the results in Figs.~\ref{fig6a}-\ref{fig6b}. In particular, we see that linear susceptibilities are close to $0$ and
only slightly different from each other when there is no SGC, so that threshold values of $U_i$ are not distinguishable for $\Omega_c=4,6,8$. However, differences in nonlinear susceptibility even in this case lead to reasonably distinct values of threshold intensity for OB. When SGC is present both the $U_i$ and $I_i$ threshold values are well distinguishable for $\Omega_c=4,6,8$.

\begin{table}
\caption{\label{table1} List of linear  $\chi^{(1)}$ and nonlinear  $\chi^{(3)}$ susceptibilities, and the corresponding probe intensity $U_i$ at the threshold of OB for probe detuning $\Delta_p=0.05$, SGC parameter $p=0.99$, and coupling field Rabi frequencies $\Omega_c^0=4,6,8$. All the parameters are scaled and dimensionless as explained in the text.}
\begin{ruledtabular}
\begin{tabular}{ccccc}
p & $\Omega_c^0$ & $\mathrm{Re}(\chi^{(1)})$ & $\mathrm{Re}(\chi^{(3)})$ & $U_i$ \\
\hline
$0$ & $4$ & $-0.0031$ & $0.00039$ & $0.31$\\
\hline
$0$ & $6$ & $-0.0014$ & $0.000077$ & $0.31$\\
\hline
$0$ & $8$ & $ -0.0008 $ & $0.000024$ & $0.31$\\
\hline
$0.99$ & $4$ & $-0.1439$ & $0.9756$ & $0.48$\\
\hline
$0.99$ & $6$ & $-0.0687$ & $0.1969$ & $0.38$\\
\hline
$0.99$ & $8$ & $-0.0391$ & $0.0621$ & $0.34$\\
\end{tabular}
\end{ruledtabular}
\end{table}

In Fig.~\ref{fig7} the effect of atomic detuning on the threshold value of OB is investigated. It can be seen that this atomic parameter can also be used for controlling of OB in the system. The threshold of OB decreases by decreasing the atomic detuning for a given $\Omega_c^0$ at the cost of lower contrast between the highest and lowest transmission outputs.  

We conclude this section with a brief discussion of implementation of our model in a real physical system. 
Observation of SGC in real atoms, despite some controversial
experimental claims, is too difficult due to the stringent requirement
of finding closely lying levels with parallel dipole moments.
More flexibility for implementing
advantages of SGC can be achieved by designing equivalent dressed state representations~\cite{wu_control_2005}. Quantum coherence of a bare state system can be mapped to SGC of dressed state picture which corresponds to quasi degenerate low lying levels required for SGC. 
On the other hand SGC has already been observed with artificial atoms such
as charged quantum dots~\cite{dutt_stimulated_2005}. Embedding
GaAs quantum dot structures in photonic crystals~\cite{kuroda_acceleration_2008} is an avaliable
techonology which can be tailored to implement 
our model system. Semiconductor quantum well heterostructures 
can also be used
in place of the defect layer in our system. Quantum coherence in the resonant tunneling can replace the SGC effect equivalently~\cite{wu_ultrafast_2005}. 
Due to the Hartmann
effect the size of the heterostructures should not influence
the speed of the OB operation. 
The overall increase of the size of the multilayer system would
be comparable to the doped defect layer as the quantum well
heterostructure consists of few thin nanometer size layers. 
Surrounding PC layers (A,B in Fig.~\ref{fig1}) are about 
$50-80$ nm each typically. 
\section{Conclusions}\label{sec:con}
We explored the effects of spontaneously generated coherence and atomic control parameters, specifically probe detuning and control field Rabi frequency, on the characteristics of optical bistability in a 1DPC doped with $ \Lambda $-type three-level atoms. It is found that OB threshold, size of the hysteresis loop, and the contrast between OB outputs can be controlled over significantly wide ranges by these parameters. We find the parameter
regimes which allow for negligible absorption and enhanced nonlinear
response simultaneously. We discussed that the proposed model system can be implemented effectively using artificial atoms such as charged quantum dot heterostructures. Doped multilayer photonic crystals with engineered quantum coherence offer more compact solutions to flexible wide range control in photonic switching applications than present alternatives based
upon coupled systems or systems with large number of layers.
\acknowledgements
S.~A. acknowledge the accommodation support by the Office of Vice President for Academic Affairs (VPAA) and hospitality of the Department of Physics of the Ko\c{c} University.

\begin{thebibliography}{48}
\expandafter\ifx\csname natexlab\endcsname\relax\def\natexlab#1{#1}\fi
\expandafter\ifx\csname bibnamefont\endcsname\relax
  \def\bibnamefont#1{#1}\fi
\expandafter\ifx\csname bibfnamefont\endcsname\relax
  \def\bibfnamefont#1{#1}\fi
\expandafter\ifx\csname citenamefont\endcsname\relax
  \def\citenamefont#1{#1}\fi
\expandafter\ifx\csname url\endcsname\relax
  \def\url#1{\texttt{#1}}\fi
\expandafter\ifx\csname urlprefix\endcsname\relax\def\urlprefix{URL }\fi
\providecommand{\bibinfo}[2]{#2}
\providecommand{\eprint}[2][]{\url{#2}}

\bibitem[{\citenamefont{Gibbs}(1985)}]{gibbs_optical_1985}
\bibinfo{author}{\bibfnamefont{H.~M.} \bibnamefont{Gibbs}},
  \emph{\bibinfo{title}{Optical Bistability: Controlling Light with Light}}
  (\bibinfo{publisher}{Academic Press}, \bibinfo{address}{Orlando},
  \bibinfo{year}{1985}).

\bibitem[{\citenamefont{Abraham and Smith}(1982)}]{abraham_optical_1982}
\bibinfo{author}{\bibfnamefont{E.}~\bibnamefont{Abraham}} \bibnamefont{and}
  \bibinfo{author}{\bibfnamefont{S.~D.} \bibnamefont{Smith}},
  \bibinfo{journal}{Rep. Prog. Phys.} \textbf{\bibinfo{volume}{45}},
  \bibinfo{pages}{815} (\bibinfo{year}{1982}).

\bibitem[{\citenamefont{Solja{\v c}i{\'c} et~al.}(2002)\citenamefont{Solja{\v
  c}i{\'c}, Ibanescu, Johnson, Fink, and Joannopoulos}}]{soljacic_optimal_2002}
\bibinfo{author}{\bibfnamefont{M.}~\bibnamefont{Solja{\v c}i{\'c}}},
  \bibinfo{author}{\bibfnamefont{M.}~\bibnamefont{Ibanescu}},
  \bibinfo{author}{\bibfnamefont{S.~G.} \bibnamefont{Johnson}},
  \bibinfo{author}{\bibfnamefont{Y.}~\bibnamefont{Fink}}, \bibnamefont{and}
  \bibinfo{author}{\bibfnamefont{J.~D.} \bibnamefont{Joannopoulos}},
  \bibinfo{journal}{Phys. Rev. E} \textbf{\bibinfo{volume}{66}},
  \bibinfo{pages}{055601} (\bibinfo{year}{2002}).

\bibitem[{\citenamefont{Barclay et~al.}(2005)\citenamefont{Barclay, Srinivasan,
  and Painter}}]{barclay_nonlinear_2005}
\bibinfo{author}{\bibfnamefont{P.}~\bibnamefont{Barclay}},
  \bibinfo{author}{\bibfnamefont{K.}~\bibnamefont{Srinivasan}},
  \bibnamefont{and} \bibinfo{author}{\bibfnamefont{O.}~\bibnamefont{Painter}},
  \bibinfo{journal}{Opt. Express} \textbf{\bibinfo{volume}{13}},
  \bibinfo{pages}{801} (\bibinfo{year}{2005}).

\bibitem[{\citenamefont{Scalora et~al.}(1994)\citenamefont{Scalora, Dowling,
  Bowden, and Bloemer}}]{Scalora_optical_1994}
\bibinfo{author}{\bibfnamefont{M.}~\bibnamefont{Scalora}},
  \bibinfo{author}{\bibfnamefont{J.~P.} \bibnamefont{Dowling}},
  \bibinfo{author}{\bibfnamefont{C.~M.} \bibnamefont{Bowden}},
  \bibnamefont{and} \bibinfo{author}{\bibfnamefont{M.~J.}
  \bibnamefont{Bloemer}}, \bibinfo{journal}{Phys. Rev. Lett.}
  \textbf{\bibinfo{volume}{73}}, \bibinfo{pages}{1368} (\bibinfo{year}{1994}).

\bibitem[{\citenamefont{Danckaert et~al.}(1991)\citenamefont{Danckaert,
  Fobelets, Veretennicoff, Vitrant, and Reinisch}}]{Danckaert_dispersive_1991}
\bibinfo{author}{\bibfnamefont{J.}~\bibnamefont{Danckaert}},
  \bibinfo{author}{\bibfnamefont{K.}~\bibnamefont{Fobelets}},
  \bibinfo{author}{\bibfnamefont{I.}~\bibnamefont{Veretennicoff}},
  \bibinfo{author}{\bibfnamefont{G.}~\bibnamefont{Vitrant}}, \bibnamefont{and}
  \bibinfo{author}{\bibfnamefont{R.}~\bibnamefont{Reinisch}},
  \bibinfo{journal}{Phys. Rev. B} \textbf{\bibinfo{volume}{44}},
  \bibinfo{pages}{8214} (\bibinfo{year}{1991}).

\bibitem[{\citenamefont{Agranovich et~al.}(1991)\citenamefont{Agranovich,
  Kiselev, and Mills}}]{Agranovich_optical_1991}
\bibinfo{author}{\bibfnamefont{V.~M.} \bibnamefont{Agranovich}},
  \bibinfo{author}{\bibfnamefont{S.~A.} \bibnamefont{Kiselev}},
  \bibnamefont{and} \bibinfo{author}{\bibfnamefont{D.~L.} \bibnamefont{Mills}},
  \bibinfo{journal}{Phys. Rev. B} \textbf{\bibinfo{volume}{44}},
  \bibinfo{pages}{10917} (\bibinfo{year}{1991}).

\bibitem[{\citenamefont{Mingaleev and
  Kivshar}(2002)}]{mingaleev_nonlinear_2002}
\bibinfo{author}{\bibfnamefont{S.~F.} \bibnamefont{Mingaleev}}
  \bibnamefont{and} \bibinfo{author}{\bibfnamefont{Y.~S.}
  \bibnamefont{Kivshar}}, \bibinfo{journal}{J. Opt. Soc. Am. B}
  \textbf{\bibinfo{volume}{19}}, \bibinfo{pages}{2241} (\bibinfo{year}{2002}).

\bibitem[{\citenamefont{Yanik et~al.}(2003{\natexlab{a}})\citenamefont{Yanik,
  Fan, and Solja{\v c}i{\'c}}}]{yanik_high-contrast_2003}
\bibinfo{author}{\bibfnamefont{M.~F.} \bibnamefont{Yanik}},
  \bibinfo{author}{\bibfnamefont{S.}~\bibnamefont{Fan}}, \bibnamefont{and}
  \bibinfo{author}{\bibfnamefont{M.}~\bibnamefont{Solja{\v c}i{\'c}}},
  \bibinfo{journal}{Appl. Phys. Lett.} \textbf{\bibinfo{volume}{83}}, \bibinfo{pages}{2739}
  (\bibinfo{year}{2003}{\natexlab{a}}).

\bibitem[{\citenamefont{Mingaleev et~al.}(2006)\citenamefont{Mingaleev,
  Miroshnichenko, Kivshar, and Busch}}]{mingaleev_all-optical_2006}
\bibinfo{author}{\bibfnamefont{S.~F.} \bibnamefont{Mingaleev}},
  \bibinfo{author}{\bibfnamefont{A.~E.} \bibnamefont{Miroshnichenko}},
  \bibinfo{author}{\bibfnamefont{Y.~S.} \bibnamefont{Kivshar}},
  \bibnamefont{and} \bibinfo{author}{\bibfnamefont{K.}~\bibnamefont{Busch}},
  \bibinfo{journal}{Phys. Rev. E} \textbf{\bibinfo{volume}{74}},
  \bibinfo{pages}{046603} (\bibinfo{year}{2006}).

\bibitem[{\citenamefont{Tocci et~al.}(1995)\citenamefont{Tocci, Bloemer,
  Scalora, Dowling, and Bowden}}]{tocci_thinfilm_1995}
\bibinfo{author}{\bibfnamefont{M.~D.} \bibnamefont{Tocci}},
  \bibinfo{author}{\bibfnamefont{M.~J.} \bibnamefont{Bloemer}},
  \bibinfo{author}{\bibfnamefont{M.}~\bibnamefont{Scalora}},
  \bibinfo{author}{\bibfnamefont{J.~P.} \bibnamefont{Dowling}},
  \bibnamefont{and} \bibinfo{author}{\bibfnamefont{C.~M.}
  \bibnamefont{Bowden}}, \bibinfo{journal}{Appl. Phys. Lett.} \textbf{\bibinfo{volume}{66}}, \bibinfo{pages}{2324}
  (\bibinfo{year}{1995}).

\bibitem[{\citenamefont{Zhao et~al.}(2006)\citenamefont{Zhao, Zhou, Guo, Hu,
  Yang, Lan, and Lin}}]{zhao_design_2006}
\bibinfo{author}{\bibfnamefont{N.-S.} \bibnamefont{Zhao}},
  \bibinfo{author}{\bibfnamefont{H.}~\bibnamefont{Zhou}},
  \bibinfo{author}{\bibfnamefont{Q.}~\bibnamefont{Guo}},
  \bibinfo{author}{\bibfnamefont{W.}~\bibnamefont{Hu}},
  \bibinfo{author}{\bibfnamefont{X.-B.} \bibnamefont{Yang}},
  \bibinfo{author}{\bibfnamefont{S.}~\bibnamefont{Lan}}, \bibnamefont{and}
  \bibinfo{author}{\bibfnamefont{X.-S.} \bibnamefont{Lin}},
  \bibinfo{journal}{J. Opt. Soc. Am. B} \textbf{\bibinfo{volume}{23}},
  \bibinfo{pages}{2434} (\bibinfo{year}{2006}).

\bibitem[{\citenamefont{Xue et~al.}(2010)\citenamefont{Xue, Jiang, and
  Chen}}]{xue_highly_2010}
\bibinfo{author}{\bibfnamefont{C.}~\bibnamefont{Xue}},
  \bibinfo{author}{\bibfnamefont{H.}~\bibnamefont{Jiang}}, \bibnamefont{and}
  \bibinfo{author}{\bibfnamefont{H.}~\bibnamefont{Chen}},
  \bibinfo{journal}{Opt. Express} \textbf{\bibinfo{volume}{18}},
  \bibinfo{pages}{7479} (\bibinfo{year}{2010}).

\bibitem[{\citenamefont{Solja?i? et~al.}(2003)\citenamefont{Solja{\v c}i{\'c}, Luo,
  Joannopoulos, and Fan}}]{soljacic_nonlinear_2003}
\bibinfo{author}{\bibfnamefont{M.}~\bibnamefont{Solja{\v c}i{\'c}}},
  \bibinfo{author}{\bibfnamefont{C.}~\bibnamefont{Luo}},
  \bibinfo{author}{\bibfnamefont{J.~D.} \bibnamefont{Joannopoulos}},
  \bibnamefont{and} \bibinfo{author}{\bibfnamefont{S.}~\bibnamefont{Fan}},
  \bibinfo{journal}{Opt. Lett.} \textbf{\bibinfo{volume}{28}},
  \bibinfo{pages}{637} (\bibinfo{year}{2003}).

\bibitem[{\citenamefont{Yanik et~al.}(2003{\natexlab{b}})\citenamefont{Yanik,
  Fan, Solja{\v c}i{\'c}, and Joannopoulos}}]{yanik_all-optical_2003}
\bibinfo{author}{\bibfnamefont{M.~F.} \bibnamefont{Yanik}},
  \bibinfo{author}{\bibfnamefont{S.}~\bibnamefont{Fan}},
  \bibinfo{author}{\bibfnamefont{M.}~\bibnamefont{Solja{\v c}i{\'c}}}, \bibnamefont{and}
  \bibinfo{author}{\bibfnamefont{J.~D.} \bibnamefont{Joannopoulos}},
  \bibinfo{journal}{Opt. Lett.} \textbf{\bibinfo{volume}{28}},
  \bibinfo{pages}{2506} (\bibinfo{year}{2003}{\natexlab{b}}).

\bibitem[{\citenamefont{Bravo-Abad et~al.}(2007)\citenamefont{Bravo-Abad,
  Rodriguez, Bermel, Johnson, Joannopoulos, and
  Solja{\v c}i{\'c}}}]{bravo-abad_enhanced_2007}
\bibinfo{author}{\bibfnamefont{J.}~\bibnamefont{Bravo-Abad}},
  \bibinfo{author}{\bibfnamefont{A.}~\bibnamefont{Rodriguez}},
  \bibinfo{author}{\bibfnamefont{P.}~\bibnamefont{Bermel}},
  \bibinfo{author}{\bibfnamefont{S.~G.} \bibnamefont{Johnson}},
  \bibinfo{author}{\bibfnamefont{J.~D.} \bibnamefont{Joannopoulos}},
  \bibnamefont{and} \bibinfo{author}{\bibfnamefont{M.}~\bibnamefont{Solja{\v c}i{\'c}}},
  \bibinfo{journal}{Opt. Express} \textbf{\bibinfo{volume}{15}},
  \bibinfo{pages}{16161} (\bibinfo{year}{2007}).

\bibitem[{\citenamefont{Guo and L{\"u}}(2009)}]{guo_controllable_2009}
\bibinfo{author}{\bibfnamefont{X.-Y.} \bibnamefont{Guo}} \bibnamefont{and}
  \bibinfo{author}{\bibfnamefont{S.-C.} \bibnamefont{L{\"u}}},
  \bibinfo{journal}{Phys. Rev. A} \textbf{\bibinfo{volume}{80}},
  \bibinfo{pages}{043826} (\bibinfo{year}{2009}).

\bibitem[{\citenamefont{John and Quang}(1996)}]{john_optical_1996}
\bibinfo{author}{\bibfnamefont{S.}~\bibnamefont{John}} \bibnamefont{and}
  \bibinfo{author}{\bibfnamefont{T.}~\bibnamefont{Quang}},
  \bibinfo{journal}{Phys. Rev. A} \textbf{\bibinfo{volume}{54}},
  \bibinfo{pages}{4479} (\bibinfo{year}{1996}).

\bibitem[{\citenamefont{John and Quang}(1997)}]{john_collective_1997}
\bibinfo{author}{\bibfnamefont{S.}~\bibnamefont{John}} \bibnamefont{and}
  \bibinfo{author}{\bibfnamefont{T.}~\bibnamefont{Quang}},
  \bibinfo{journal}{Phys. Rev. Lett.} \textbf{\bibinfo{volume}{78}},
  \bibinfo{pages}{1888} (\bibinfo{year}{1997}).

\bibitem[{\citenamefont{Ma and John}(2011)}]{ma_optical_2011}
\bibinfo{author}{\bibfnamefont{X.}~\bibnamefont{Ma}} \bibnamefont{and}
  \bibinfo{author}{\bibfnamefont{S.}~\bibnamefont{John}},
  \bibinfo{journal}{Phys. Rev. A} \textbf{\bibinfo{volume}{84}},
  \bibinfo{pages}{053848} (\bibinfo{year}{2011}).

\bibitem[{\citenamefont{Takeda and John}(2011)}]{takeda_self-consistent_2011}
\bibinfo{author}{\bibfnamefont{H.}~\bibnamefont{Takeda}} \bibnamefont{and}
  \bibinfo{author}{\bibfnamefont{S.}~\bibnamefont{John}},
  \bibinfo{journal}{Phys. Rev. A} \textbf{\bibinfo{volume}{83}},
  \bibinfo{pages}{053811} (\bibinfo{year}{2011}).

\bibitem[{\citenamefont{Vujic and John}(2007)}]{vujic_coherent_2007}
\bibinfo{author}{\bibfnamefont{D.}~\bibnamefont{Vujic}} \bibnamefont{and}
  \bibinfo{author}{\bibfnamefont{S.}~\bibnamefont{John}},
  \bibinfo{journal}{Phys. Rev. A} \textbf{\bibinfo{volume}{76}},
  \bibinfo{pages}{063814} (\bibinfo{year}{2007}).

\bibitem[{\citenamefont{Wang et~al.}(1997)\citenamefont{Wang, Dong, and
  Xing}}]{wang_dispersive_1997}
\bibinfo{author}{\bibfnamefont{R.}~\bibnamefont{Wang}},
  \bibinfo{author}{\bibfnamefont{J.}~\bibnamefont{Dong}}, \bibnamefont{and}
  \bibinfo{author}{\bibfnamefont{D.~Y.} \bibnamefont{Xing}},
  \bibinfo{journal}{Phys. Rev. E} \textbf{\bibinfo{volume}{55}},
  \bibinfo{pages}{6301} (\bibinfo{year}{1997}).

\bibitem[{\citenamefont{Lidorikis et~al.}(1997)\citenamefont{Lidorikis, Busch,
  Li, Chan, and Soukoulis}}]{Lidorikis_optical_1997}
\bibinfo{author}{\bibfnamefont{E.}~\bibnamefont{Lidorikis}},
  \bibinfo{author}{\bibfnamefont{K.}~\bibnamefont{Busch}},
  \bibinfo{author}{\bibfnamefont{Q.}~\bibnamefont{Li}},
  \bibinfo{author}{\bibfnamefont{C.~T.} \bibnamefont{Chan}}, \bibnamefont{and}
  \bibinfo{author}{\bibfnamefont{C.~M.} \bibnamefont{Soukoulis}},
  \bibinfo{journal}{Phys. Rev. B} \textbf{\bibinfo{volume}{56}},
  \bibinfo{pages}{15090} (\bibinfo{year}{1997}).

\bibitem[{\citenamefont{Novitsky and
  Mikhnevich}(2008)}]{novitsky_bistable_2008}
\bibinfo{author}{\bibfnamefont{D.~V.} \bibnamefont{Novitsky}} \bibnamefont{and}
  \bibinfo{author}{\bibfnamefont{S.~Y.} \bibnamefont{Mikhnevich}},
  \bibinfo{journal}{J. Opt. Soc. Am. B} \textbf{\bibinfo{volume}{25}},
  \bibinfo{pages}{1362} (\bibinfo{year}{2008}).
  
    
    \bibitem[{\citenamefont{Gupta and Agarwal}(1987)}]{gupta_dispersive_1987}
    \bibinfo{author}{\bibfnamefont{S.~D.} \bibnamefont{Gupta}} \bibnamefont{and}
      \bibinfo{author}{\bibfnamefont{G.~S.} \bibnamefont{Agarwal}},
      \bibinfo{journal}{Journal of the Optical Society of America B}
      \textbf{\bibinfo{volume}{4}}, \bibinfo{pages}{691} (\bibinfo{year}{1987}).

\bibitem[{\citenamefont{He and Cada}(1992)}]{he_combined_1992}
\bibinfo{author}{\bibfnamefont{J.}~\bibnamefont{He}} \bibnamefont{and}
  \bibinfo{author}{\bibfnamefont{M.}~\bibnamefont{Cada}},
  \bibinfo{journal}{Appl. Phys. Lett.} \textbf{\bibinfo{volume}{61}}, \bibinfo{pages}{2150} (\bibinfo{year}{1992}).

\bibitem[{\citenamefont{Jose}(2009)}]{jose_controlling_2009}
\bibinfo{author}{\bibfnamefont{J.}~\bibnamefont{Jose}}, \bibinfo{journal}{J.
  Phys. B: At. Mol. Opt. Phys.} \textbf{\bibinfo{volume}{42}},
  \bibinfo{pages}{095401} (\bibinfo{year}{2009}).

\bibitem[{\citenamefont{Hou et~al.}(2008)\citenamefont{Hou, Chen, Shi, Kong,
  Ge, and Wang}}]{hou_transmission_2008}
\bibinfo{author}{\bibfnamefont{P.}~\bibnamefont{Hou}},
  \bibinfo{author}{\bibfnamefont{Y.}~\bibnamefont{Chen}},
  \bibinfo{author}{\bibfnamefont{J.}~\bibnamefont{Shi}},
  \bibinfo{author}{\bibfnamefont{Q.}~\bibnamefont{Kong}},
  \bibinfo{author}{\bibfnamefont{L.}~\bibnamefont{Ge}}, \bibnamefont{and}
  \bibinfo{author}{\bibfnamefont{Q.}~\bibnamefont{Wang}},
  \bibinfo{journal}{Appl. Phys. A - Mater} \textbf{\bibinfo{volume}{91}}, \bibinfo{pages}{41} (\bibinfo{year}{2008}).

\bibitem[{\citenamefont{Walls and Zoller}(1980)}]{walls_coherent_1980}
\bibinfo{author}{\bibfnamefont{D.}~\bibnamefont{Walls}} \bibnamefont{and}
  \bibinfo{author}{\bibfnamefont{P.}~\bibnamefont{Zoller}},
  \bibinfo{journal}{Optics Communications} \textbf{\bibinfo{volume}{34}},
  \bibinfo{pages}{260} (\bibinfo{year}{1980}).

\bibitem[{\citenamefont{Walls et~al.}(1981)\citenamefont{Walls, Zoller, and
  Steyn-Ross}}]{Walls_optical_1981}
\bibinfo{author}{\bibfnamefont{D.}~\bibnamefont{Walls}},
  \bibinfo{author}{\bibfnamefont{P.}~\bibnamefont{Zoller}}, \bibnamefont{and}
  \bibinfo{author}{\bibfnamefont{M.}~\bibnamefont{Steyn-Ross}},
  \bibinfo{journal}{IEEE J. Quant. Electron} \textbf{\bibinfo{volume}{17}}, \bibinfo{pages}{380} (\bibinfo{year}{1981}).

\bibitem[{\citenamefont{Harshawardhan and
  Agarwal}(1996)}]{Harshawardhan_controlling_1996}
\bibinfo{author}{\bibfnamefont{W.}~\bibnamefont{Harshawardhan}}
  \bibnamefont{and} \bibinfo{author}{\bibfnamefont{G.~S.}
  \bibnamefont{Agarwal}}, \bibinfo{journal}{Phys. Rev. A} \textbf{\bibinfo{volume}{53}}, \bibinfo{pages}{1812}
  (\bibinfo{year}{1996}).

\bibitem[{\citenamefont{Ant{\'o}n and Calder{\'o}n}(2002)}]{anton_optical_2002}
\bibinfo{author}{\bibfnamefont{M.~A.} \bibnamefont{Ant{\'o}n}}
  \bibnamefont{and} \bibinfo{author}{\bibfnamefont{O.~G.}
  \bibnamefont{Calder{\'o}n}}, \bibinfo{journal}{J Opt. B-Quantum S. O.} \textbf{\bibinfo{volume}{4}},
  \bibinfo{pages}{91} (\bibinfo{year}{2002}).

\bibitem[{\citenamefont{Ant{\'o}n et~al.}(2003)\citenamefont{Ant{\'o}n,
  Calder{\'o}n, and Carre{\~n}o}}]{anton_optical_2003}
\bibinfo{author}{\bibfnamefont{M.}~\bibnamefont{Ant{\'o}n}},
  \bibinfo{author}{\bibfnamefont{O.~G.} \bibnamefont{Calder{\'o}n}},
  \bibnamefont{and}
  \bibinfo{author}{\bibfnamefont{F.}~\bibnamefont{Carre{\~n}o}},
  \bibinfo{journal}{Physics Letters A} \textbf{\bibinfo{volume}{311}},
  \bibinfo{pages}{297} (\bibinfo{year}{2003}).

\bibitem[{\citenamefont{Wang and Xu}(2009)}]{wang_control_2009}
\bibinfo{author}{\bibfnamefont{Z.}~\bibnamefont{Wang}} \bibnamefont{and}
  \bibinfo{author}{\bibfnamefont{M.}~\bibnamefont{Xu}},
  \bibinfo{journal}{Optics Communications} \textbf{\bibinfo{volume}{282}},
  \bibinfo{pages}{1574} (\bibinfo{year}{2009}).

\bibitem[{\citenamefont{Javanainen}(1992)}]{javanainen_effect_1992}
\bibinfo{author}{\bibfnamefont{J.}~\bibnamefont{Javanainen}},
  \bibinfo{journal}{{EPL}} \textbf{\bibinfo{volume}{17}}, \bibinfo{pages}{407}
  (\bibinfo{year}{1992}).

\bibitem[{\citenamefont{Niu and Gong}(2006)}]{niu_enhancing_2006}
\bibinfo{author}{\bibfnamefont{Y.}~\bibnamefont{Niu}} \bibnamefont{and}
  \bibinfo{author}{\bibfnamefont{S.}~\bibnamefont{Gong}},
  \bibinfo{journal}{Phys. Rev. A} \textbf{\bibinfo{volume}{73}},
  \bibinfo{pages}{053811} (\bibinfo{year}{2006}).

\bibitem[{\citenamefont{Braun et~al.}(2006)\citenamefont{Braun, Rinne, and
  Garc{\'\i}a-Santamar{\'\i}a}}]{braun_introducing_2006}
\bibinfo{author}{\bibfnamefont{P.~V.} \bibnamefont{Braun}},
  \bibinfo{author}{\bibfnamefont{S.~A.} \bibnamefont{Rinne}}, \bibnamefont{and}
  \bibinfo{author}{\bibfnamefont{F.}~\bibnamefont{Garc{\'\i}a-Santamar{\'\i}a}}, \bibinfo{journal}{Adv. Mater.}
  \textbf{\bibinfo{volume}{18}}, \bibinfo{pages}{2665} (\bibinfo{year}{2006}).
  
  \bibitem[{\citenamefont{Kuroda et~al.}(2008)\citenamefont{Kuroda, Ikeda, Mano,
    Sugimoto, Ochiai, Kuroda, Ohkouchi, Koguchi, Sakoda, and
    Asakawa}}]{kuroda_acceleration_2008}
  \bibinfo{author}{\bibfnamefont{T.}~\bibnamefont{Kuroda}},
    \bibinfo{author}{\bibfnamefont{N.}~\bibnamefont{Ikeda}},
    \bibinfo{author}{\bibfnamefont{T.}~\bibnamefont{Mano}},
    \bibinfo{author}{\bibfnamefont{Y.}~\bibnamefont{Sugimoto}},
    \bibinfo{author}{\bibfnamefont{T.}~\bibnamefont{Ochiai}},
    \bibinfo{author}{\bibfnamefont{K.}~\bibnamefont{Kuroda}},
    \bibinfo{author}{\bibfnamefont{S.}~\bibnamefont{Ohkouchi}},
    \bibinfo{author}{\bibfnamefont{N.}~\bibnamefont{Koguchi}},
    \bibinfo{author}{\bibfnamefont{K.}~\bibnamefont{Sakoda}}, \bibnamefont{and}
    \bibinfo{author}{\bibfnamefont{K.}~\bibnamefont{Asakawa}},
    \bibinfo{journal}{Appl. Phys. Lett.} \textbf{\bibinfo{volume}{93}},
    \bibinfo{pages}{111103} (\bibinfo{year}{2008}).
  
\bibitem[{\citenamefont{Dutt et~al.}(2005)\citenamefont{Dutt, Cheng, Li, Xu,
  Li, Berman, Steel, Bracker, Gammon, Economou et~al.}}]{dutt_stimulated_2005}
\bibinfo{author}{\bibfnamefont{M.~V.~G.} \bibnamefont{Dutt}},
  \bibinfo{author}{\bibfnamefont{J.}~\bibnamefont{Cheng}},
  \bibinfo{author}{\bibfnamefont{B.}~\bibnamefont{Li}},
  \bibinfo{author}{\bibfnamefont{X.}~\bibnamefont{Xu}},
  \bibinfo{author}{\bibfnamefont{X.}~\bibnamefont{Li}},
  \bibinfo{author}{\bibfnamefont{P.~R.} \bibnamefont{Berman}},
  \bibinfo{author}{\bibfnamefont{D.~G.} \bibnamefont{Steel}},
  \bibinfo{author}{\bibfnamefont{A.~S.} \bibnamefont{Bracker}},
  \bibinfo{author}{\bibfnamefont{D.}~\bibnamefont{Gammon}},
  \bibinfo{author}{\bibfnamefont{S.~E.} \bibnamefont{Economou}},
  \bibnamefont{et~al.}, \bibinfo{journal}{Phys. Rev. Lett.}
  \textbf{\bibinfo{volume}{94}}, \bibinfo{pages}{227403}
  (\bibinfo{year}{2005}).

\bibitem[{\citenamefont{Wu et~al.}(2005{\natexlab{b}})\citenamefont{Wu, Gao,
  Xu, Silvestri, Artoni, La~Rocca, and Bassani}}]{wu_ultrafast_2005}
\bibinfo{author}{\bibfnamefont{J.-H.} \bibnamefont{Wu}},
  \bibinfo{author}{\bibfnamefont{J.-Y.} \bibnamefont{Gao}},
  \bibinfo{author}{\bibfnamefont{J.-H.} \bibnamefont{Xu}},
  \bibinfo{author}{\bibfnamefont{L.}~\bibnamefont{Silvestri}},
  \bibinfo{author}{\bibfnamefont{M.}~\bibnamefont{Artoni}},
  \bibinfo{author}{\bibfnamefont{G.~C.} \bibnamefont{La~Rocca}},
  \bibnamefont{and} \bibinfo{author}{\bibfnamefont{F.}~\bibnamefont{Bassani}},
  \bibinfo{journal}{Phys. Rev. Lett.} \textbf{\bibinfo{volume}{95}},
  \bibinfo{pages}{057401} (\bibinfo{year}{2005}{\natexlab{b}}).
      
  \bibitem[{\citenamefont{Wu et~al.}(2005{\natexlab{a}})\citenamefont{Wu, Li,
    Ding, Zhao, and Gao}}]{wu_control_2005}
  \bibinfo{author}{\bibfnamefont{J.-H.} \bibnamefont{Wu}},
    \bibinfo{author}{\bibfnamefont{A.-J.} \bibnamefont{Li}},
    \bibinfo{author}{\bibfnamefont{Y.}~\bibnamefont{Ding}},
    \bibinfo{author}{\bibfnamefont{Y.-C.} \bibnamefont{Zhao}}, \bibnamefont{and}
    \bibinfo{author}{\bibfnamefont{J.-Y.} \bibnamefont{Gao}},
    \bibinfo{journal}{Phys. Rev. A} \textbf{\bibinfo{volume}{72}},
    \bibinfo{pages}{023802} (\bibinfo{year}{2005}{\natexlab{a}}).

\bibitem[{\citenamefont{Jiang et~al.}(2003)\citenamefont{Jiang, Chen, Li,
  Zhang, and Zhu}}]{jiang_omnidirectional_2003}
\bibinfo{author}{\bibfnamefont{H.}~\bibnamefont{Jiang}},
  \bibinfo{author}{\bibfnamefont{H.}~\bibnamefont{Chen}},
  \bibinfo{author}{\bibfnamefont{H.}~\bibnamefont{Li}},
  \bibinfo{author}{\bibfnamefont{Y.}~\bibnamefont{Zhang}}, \bibnamefont{and}
  \bibinfo{author}{\bibfnamefont{S.}~\bibnamefont{Zhu}},
  \bibinfo{journal}{Appl. Phys. Lett.} \textbf{\bibinfo{volume}{83}}, \bibinfo{pages}{5386} (\bibinfo{year}{2003}).

\bibitem[{\citenamefont{Steinberg et~al.}(1993)\citenamefont{Steinberg, Kwiat,
  and Chiao}}]{steinberg_measurement_1993}
\bibinfo{author}{\bibfnamefont{A.~M.} \bibnamefont{Steinberg}},
  \bibinfo{author}{\bibfnamefont{P.~G.} \bibnamefont{Kwiat}}, \bibnamefont{and}
  \bibinfo{author}{\bibfnamefont{R.~Y.} \bibnamefont{Chiao}},
  \bibinfo{journal}{Phys. Rev. Lett.} \textbf{\bibinfo{volume}{71}},
  \bibinfo{pages}{708} (\bibinfo{year}{1993}).

\bibitem[{\citenamefont{Agarwal and Gupta}(1987)}]{agarwal_effect_1987}
\bibinfo{author}{\bibfnamefont{G.~S.} \bibnamefont{Agarwal}} \bibnamefont{and}
  \bibinfo{author}{\bibfnamefont{S.~D.} \bibnamefont{Gupta}},
  \bibinfo{journal}{Opt. Lett.} \textbf{\bibinfo{volume}{12}},
  \bibinfo{pages}{829} (\bibinfo{year}{1987}).

\bibitem[{\citenamefont{Gupta and Ray}(1988)}]{gupta_optical_1988}
\bibinfo{author}{\bibfnamefont{S.~D.} \bibnamefont{Gupta}} \bibnamefont{and}
  \bibinfo{author}{\bibfnamefont{D.~S.} \bibnamefont{Ray}},
  \bibinfo{journal}{Phys. Rev. B} \textbf{\bibinfo{volume}{38}},
  \bibinfo{pages}{3628} (\bibinfo{year}{1988}).

\bibitem[{\citenamefont{Gupta}(1989)}]{gupta_optical_1989}
\bibinfo{author}{\bibfnamefont{S.~D.} \bibnamefont{Gupta}},
  \bibinfo{journal}{J. Opt. Soc. Am. B} \textbf{\bibinfo{volume}{6}},
  \bibinfo{pages}{1927} (\bibinfo{year}{1989}).
  
  \bibitem[{\citenamefont{Sahrai and Esfahlani}(2013)}]{sahrai_hartman_2013}
  \bibinfo{author}{\bibfnamefont{M.}~\bibnamefont{Sahrai}} \bibnamefont{and}
    \bibinfo{author}{\bibfnamefont{B.}~\bibnamefont{Esfahlani}},
    \bibinfo{journal}{Physica E}
    \textbf{\bibinfo{volume}{47}}, \bibinfo{pages}{66} (\bibinfo{year}{2013}).
  
  \bibitem[{\citenamefont{{MacColl}}(1932)}]{maccoll_note_1932}
  \bibinfo{author}{\bibfnamefont{L.~A.} \bibnamefont{{MacColl}}},
    \bibinfo{journal}{Phys. Rev.} \textbf{\bibinfo{volume}{40}},
    \bibinfo{pages}{621} (\bibinfo{year}{1932}).
  
  \bibitem[{\citenamefont{Hartman}(1962)}]{hartman_tunneling_1962}
  \bibinfo{author}{\bibfnamefont{T.~E.} \bibnamefont{Hartman}},
    \bibinfo{journal}{J. Appl. Phys.} \textbf{\bibinfo{volume}{33}}, \bibinfo{pages}{3427} (\bibinfo{year}{1962}).

\bibitem[{\citenamefont{Agarwal and Gupta}(1986)}]{agarwal_exact_1986}
\bibinfo{author}{\bibfnamefont{G.~S.} \bibnamefont{Agarwal}} \bibnamefont{and}
  \bibinfo{author}{\bibfnamefont{S.~D.} \bibnamefont{Gupta}},
  \bibinfo{journal}{Phys. Rev. B} \textbf{\bibinfo{volume}{34}},
  \bibinfo{pages}{5239} (\bibinfo{year}{1986}).


\end{thebibliography}

\end{document}